# All-dielectric high-Q dynamically tunable transmissive metasurfaces

*Ruzan Sokhoyan[1], Claudio U. Hail[1], Morgan Foley[2], Meir Y. Grajower[1], and Harry A. Atwater[1*]*

[1] Thomas J. Watson Laboratories of Applied Physics, California Institute of Technology, Pasadena, California 91125, USA

[2] Department of Physics, California Institute of Technology, Pasadena, California 911125

**Abstract:** Active metasurfaces, which are arrays of actively tunable resonant elements, can dynamically control the wavefront of the scattered light at a subwavelength scale. To date, most active metasurfaces that enable dynamic wavefront shaping operate in reflection. On the other hand, active metasurfaces operating in transmission are of considerable interest as they can readily be integrated with chip-scale light sources, yielding ultra-compact wavefront shaping devices. Here, we report designs for all-dielectric low-loss active metasurfaces which can dynamically manipulate the transmitted light wavefront in the near-infrared wavelength range. Our active metasurfaces feature an array of amorphous silicon (a-Si) pillars on a silica ($SiO_2$) substate, which support resonances with quality factors (Q-factors) as high as 9800, as well as other lower-Q resonances. First, we demonstrate that high-Q resonance dips observed in transmission can be transformed into a transmission resonance peak by positioning a-Si pillar resonators at a prescribed distance from a crystalline Si substrate, defined by an $SiO_2$ spacer layer. Next, we report the design of metasurface geometry with realistic interconnect architectures that enable thermo-optic dynamic beam switching with switching times as low as 7.3 µs. Beam switching is observed for refractive index differences between neighboring metasurface elements as low as 0.0026. Finally, we demonstrate that metasurface structures with both high-Q and lower-Q modes and realistic interconnect architectures can be used for dynamic beam steering.

**Introduction**

The prospect of creating chip-scale ultracompact optical components that can be dynamically programmed to scatter or emit light with an arbitrarily shaped wavefront has long captivated researchers. Realization of chip-scale dynamically programmable optical components operating in the near-infrared or visible wavelength range could be instrumental for important applications, such as light detection and ranging (LiDAR), free space optical communications, additive manufacturing, and directed energy. The quest for these chip-scale spatial light modulators has motivated intensive research efforts in the design of active metasurfaces[1-4]. A prototypical active metasurface is composed of an array of geometrically identical subwavelength resonant metasurface elements which can dynamically control the phase and amplitude of light scattered by each metasurface element[5]. Recent reports[1,6] have experimentally demonstrated programmable metasurface chips with electric field effect control of the phase of light scattered by each metasurface element. Dynamic control of the phase at a subwavelength scale has enabled the experimental realization of beam steering and reconfigurable focusing using the same metasurface structure[1]. Notably, this demonstration[1] and the majority of other experimentally reported active metasurfaces operate in reflection. Operation in reflection mode dictates that the illuminating light source is located off-chip, which would unavoidably increase the form factor of the resulting optical system. Moreover, a reflection configuration for the illuminating light source may limit versatility, since part of the metasurface aperture will be blocked by the light source. Transmissive metasurfaces, on the other hand, have the potential to yield more compact monolithic optical systems, since they allow for integration with chip-based light sources such as vertical cavity surface emitting lasers (VCSELs)[7] or photonic cavity surface emitting lasers (PCSELs)[8].

Dynamic amplitude-tunable transmissive metasurfaces have been experimentally demonstrated using ITO field effect modulation[9,10], conductive polymer electrochemical transitions[11], ionic transport[12], Pockels effect in organic molecules[13], and thermo-optic effect in silicon (Si)[14]. Several researchers [10,11] have utilized transmittance modulation to achieve diffractive beam switching with modest efficiencies. However dynamically tunable phase control is also a prerequisite for versatile wavefront manipulation. Recent experiments have demonstrated dynamic beam switching in transmission using reorientation of liquid crystals in which a degree of phase control has contributed to the observed dynamic beam switching[2]. Previous works have also used dielectric elastomer actuators[15] and microelectromechanical systems[16] to demonstrate adaptive metalenses. These works[15,16], however, do not allow for arbitrary wavefront reconfiguration of the transmitted light.

Several designs for dynamically tunable transmissive metasurfaces have been proposed using field effect control of transmission phase[17,18] to dynamically shape the transmitted light wavefront but reported low optical efficiencies (<0.07%), limiting their use in practical applications. Recent theoretical work[19] has also demonstrated an all-dielectric transmissive metasurface which uses carrier injection in Si to realize subwavelength phase control in transmission as well as dynamic wavefront shaping. The optical efficiency of the reported metasurface is ~65%. The maximal phase shift reported for one-dimensional phase gradients was 215°, for an assumed refractive index of silicon $\Delta n$ ~ 0.01[19]. Inverse design offers the prospect for high-directivity beam steering using all-dielectric transmissive metasurfaces, such as those based on reorientation of liquid crystal molecules[20]. It has been shown[21] that the optical anisotropy of the active liquid crystal medium surrounding the metasurface can be used to achieve large phase modulation in transmission while maintaining an optical efficiency of ~100%. Prior research[22] has also used the concept of congener dipoles to develop a dynamic phase-change metasurface design that enables transmissive phase modulation covering over 240° while maintaining transmittance exceeding 80%. These recent theoretical studies[21,22], however, do not assess wavefront shaping capabilities of the designed high-efficiency metasurfaces. Prior research[23] has also reported a thermo-optically reconfigurable metalens, which enables a continuous tunability of its focal length from 165 µm to 135 µm when the metalens temperature is increased from 20 °C to 260 °C. In our work, we use high quality factor subwavelength resonators as metasurface building blocks that enables achieving dynamically tunable optical response upon modest modulation of the external stimulus.

In the last few years, all-dielectric passive metasurfaces exhibiting high quality factors have been explored by researchers[24]. All-dielectric metasurfaces supporting delocalized photonic bound states in the continuum (BICs) have been demonstrated to show narrow-bandwidth resonances where a large electric field enhancement is observed[24,25]. In addition to structures that support delocalized modes, individual subwavelength dielectric resonators can support so-called quasi-BIC modes, also referred to as supercavity modes[26,27], which exhibit moderately high quality factors and are weakly coupled to the radiative continuum. The high quality factor of the supercavity modes originates from interference of multiple localized modes supported by a resonator[28]. These distinct features have enabled use of quasi-BIC metasurfaces for numerous applications such as sensing[29] and harmonic generation[30]. Quasi-BIC mode subwavelength nanolasers have also been realized[31]. Quasi-BIC modes supported by an individual cylinder, however, cannot be efficiently excited by a normally incident linearly polarized light, and azimuthally polarized excitation is required[26,27]. For transmissive metasurfaces, normal incidence illumination with linearly polarized light is important for metasurface integration with chip-scale light sources. Notably, addition of an appropriately spaced back reflector to a cylinder array enables excitation of array quasi-BIC modes with normally incident light[32], but a back reflector precludes use for transmissive metasurfaces. Excitation of quasi-BIC (or, equivalently, supercavity) modes by a normally incident linearly polarized plane wave using a single high-index rectangular parallelepiped has also been reported[33].

Here, we develop all-dielectric high-Q metasurface designs for dynamic beam switching or beam steering in transmission mode, utilizing the modest refractive index modulation achievable by thermo-optic modulation of amorphous Si (a-Si) with assumed index modulation ranging between $\Delta n$ = 0.0026 to $\Delta n$ = 0.01. Prior research[34] has theoretically demonstrated a high-Q beam steering active metasurface, which utilizes lithium niobate as an active material. The designed metasurface[34], however, operates in reflection. Our *transmissive* active metasurface operates at near infrared wavelengths and can be excited by a normally incident linearly polarized light, exploiting either a high-Q mode or lower-Q modes supported by an individual a-Si rectangular parallelepiped (hereafter, for brevity, referred to as a square pillar). The spectral line shape of the high-Q mode at resonance can be 'inverted' from exhibiting a dip in transmission to exhibiting a transmission peak by including a Si substrate separated from the square pillar by a silica ($SiO_2$) layer of appropriately chosen thickness. Notably, the resonances discussed in our work also exhibit large (~300°) transmitted light phase spectral variation near resonance, which is a prerequisite for dynamically tunable phase shifts. We explore how refractive index modulation of modes of individual a-Si pillars can sculp the transmitted light wavefront both in the near and far field. Finally, we report designs for physically realizable interconnect architectures to enable dynamic beam steering via thermo-optic modulation.

**Results and Discussion**

Our metasurface motif consists of an array of a-Si rectangular pillars on a silica ($SiO_2$) substrate (Figs. 1a and 1b). We consider metasurface illumination by a linearly polarized (*x*-polarized) plane wave from within the $SiO_2$ substrate, and study phase and amplitude characteristics of the transmitted light. We find that this metasurface unit cell supports a high-Q optical mode for geometrical parameters *h* = 860 nm and *l* = *w* = 963 nm (see Figs. 1b and 1c), and a metasurface period of $P_x$ = $P_y$ = 1425 nm. The refractive indices of a-Si and $SiO_2$ are taken as 3.734 [35] and 1.44, respectively. The Q-factor of the supported mode is Q = 9800, as determined by fitting the transmittance spectrum with Fano form lineshape (Supporting Information, Part 1). The observed transmittance dip is accompanied by a broad spectral feature in the transmitted light phase (Fig. 1c), indicating that this unit cell motif can permit design of metasurface phase gradients either via geometric tuning or an external active control.

To gain further insight about the high-Q mode, we investigate its spatial mode profile (see Figs. 1d and 1e and Supporting Information, Part 2). Figure 1d indicates that in the *x*-direction, the electric field is tightly confined within the resonator and is enhanced by a factor of almost 80. While the largest electric field enhancement is inside the a-Si resonator, we also observe a more modest enhancement of the electric field below and above the a-Si pillar. On the other hand, in the *y-z* plane, which is perpendicular only non-zero component is E$_x$ (Fig. 1e). In the $y$ direction, a non-negligible electric field enhancement is observed between neighboring metasurface unit cells, which could be an indication of significant inter-resonator near-field coupling in the direction perpendicular to the incoming electric field. We find that this inter-resonator coupling has broad implications for comprehensive transmitted light wavefront control.

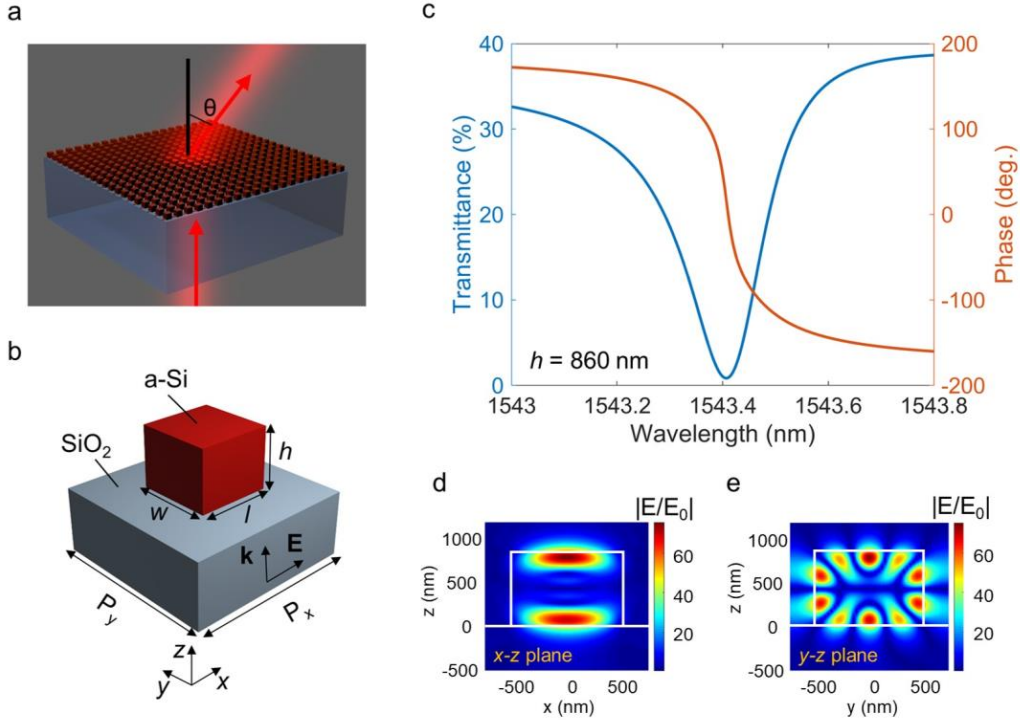

**Fig. 1.** All-dielectric high-Q metasurfaces. a) Schematic of a transmissive metasurface, which consists of an array of amorphous a-Si rectangular pillars on an SiO$_2$ substrate. b) Schematic of a unit cell of a transmissive metasurface. c) Transmittance and phase spectra of the metasurface depicted in a). In c), the assumed geometrical parameters are as follows: the pillar height is $h$ = 860 nm, and the pillar length $l$ and width $w$ are $l$ = $w$ = 963 nm. The metasurface period is $P_x$ = $P_y$ = 1425 nm. The a-Si resonators are illuminated by an $x$-polarized light from within the substrate. d) and e) show spatial distribution of the electric field amplitude in the metasurface unit cell in $x$-$z$ and $y$-$z$ planes, respectively. E$_0$ denotes the amplitude of the impinging electric field. Both in d) and e), the considered cross-section in which the electric field is calculated traverses the center of the a-Si resonator.

As a next step, we explore how the transmittance characteristics vary with the height of the square a-Si pillars while the width is kept fixed (Fig. 2a). We observe that both the resonance lineshape and linewidth change upon increasing the pillar height (Fig. 2b and 2c). At certain pillar heights, we observe asymmetric resonance lineshapes indicative of Fano resonances, namely a high-Q mode coupled to a broader mode or a "continuum" of modes. By fitting transmittance spectra to Fano resonance lineshapes, we obtained the dependence of the resonance quality factor and the Fano phase on the pillar height (see Supporting Information, Part 1). Notably, the Fano phase characterizes the phase difference between the narrow and broad resonant modes. We observe that the quality factor of the resonance increases with the pillar height, and peaks at $h$ = 860 nm (Fig. 2b), with a value of approximately 9800. At a pillar height corresponding to the maximal Q-factor, the Fano phase passes through 90° (thus, the Fano asymmetry parameter is zero), indicating a symmetric quasi-Lorentzian lineshape. Interestingly, if we remove the substrate and consider an array of a-Si pillars suspended in free space, Q-factor can be as high as 221,000 (see Supporting Information, Part 3). It may be possible to achieve even higher quality factors in free space by even more systematic variation of the pillar height. When changing the metasurface period, we observe that for periods exceeding 1300 nm, the spectral position of the resonance does not change significantly when increasing the metasurface period (Supporting Information, Part 4).

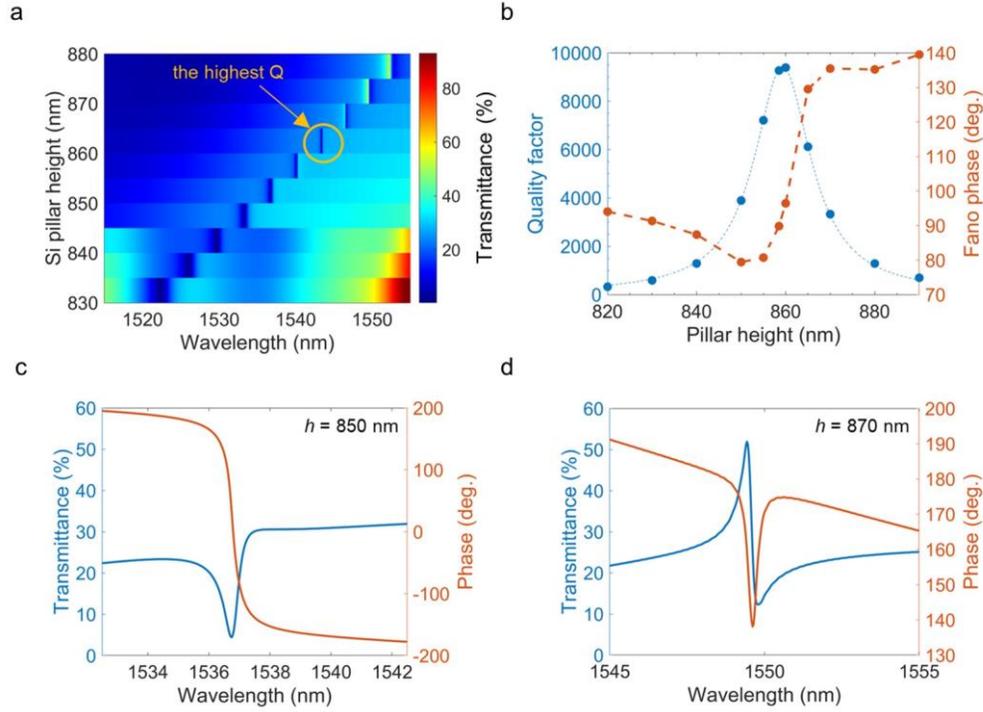

**Fig.2.** Dependence of the properties of the high-Q mode on the height of the a-Si square pillar. Transmittance a) as a function of wavelength and the a-Si pillar height $h$. b) Quality factor and Fano phase of the designed high-Q mode a function of the a-Si pillar height $h$. c) and d) show transmittance and phase as a function of wavelength for the a-Si pillar heights of $h$ = 850 nm and $h$ = 870 nm, respectively.

We also examine the spectral behavior of the transmitted light phase, which is of importance for metasurface applications. When the pillar height is below 860 nm ($h$ ≤ 860 nm), we observe that the phase of the transmitted light spans almost 360° when varying the wavelength of the transmitted light. On the other hand, when $h$ > 860 nm, we observe an abrupt change in the spectral characteristics of the transmitted light (see Supporting Information, Part 5). For example, at a pillar height of $h$ = 870 nm, the phase spans only 40° (as a function of wavelength). Interestingly, for $h$ > 860 nm, the phase variation is limited despite a high quality factor. Thus, we find that the pillar height should be below 860 nm for wavefront shaping applications.

We compare the optical mode supported by a pillar in an array to the mode of a single isolated pillar (Supporting Information, Part 5). An isolated subwavelength pillar on an $SiO_2$ substrate supports a high-Q mode with a Q-factor of 676, and the high-Q mode profile is identical to that observed in the array configuration (Fig. 1e). For an isolated a-Si pillar in free space, the Q-factor of this mode exceeds 1000, whereas, for an a-Si resonator on a substrate, the Q-factor is reduced due to mode leakage into the substrate. Scattering cross-section calculations indicate that, besides the high-Q mode, the isolated Si pillar also supports two lower-Q modes (Supporting Information, Part 5). When comparing isolated pillar mode profiles with mode profiles of a pillar in an array, we observe evidence of nonlocality for one of the lower-Q modes since its field profile is not observed in the isolated resonator (cf. Supporting Information, Part 5 and Part 3).

One shortcoming of this metasurface motif is that large phase variation is accompanied by low transmittance, and operating at the transmittance dip wavelength would result in low metasurface optical efficiency. A design modification that enables high Q-factor modes to be supported with large phase variation at a transmittance peak rather than a transmittance dip would be highly desirable. We find that the metasurface unit cell design shown in Fig. 3a, with a crystalline silicon (c-Si) substrate and a SiO$_2$ spacer, enables these criteria to be achieved. The length and width of the a-Si pillar are as in our original design ($l$ = $w$ = 963 nm), but the height of the a-Si resonator has been slightly reduced to $h$ = 845 nm, since considering taller pillars compromises variation of phase as a function of wavelength. The metasurface period is now chosen as $P_x$ = 1520 nm and $P_y$ = 1425 nm. Increasing the metasurface period in the $x$-direction to $P_x$ = 1520 nm reduces the near-field coupling between the neighboring metasurface unit cells. The spectral shape of the transmittance can be controlled by changing the thickness of the SiO$_2$ spacer. We find, for example, that an SiO$_2$ spacer thickness of $d$ = 1450 nm yields a large phase variation with wavelength accompanied by a transmittance peak of 37% (Fig. 3b and Supporting Information, Part 6).

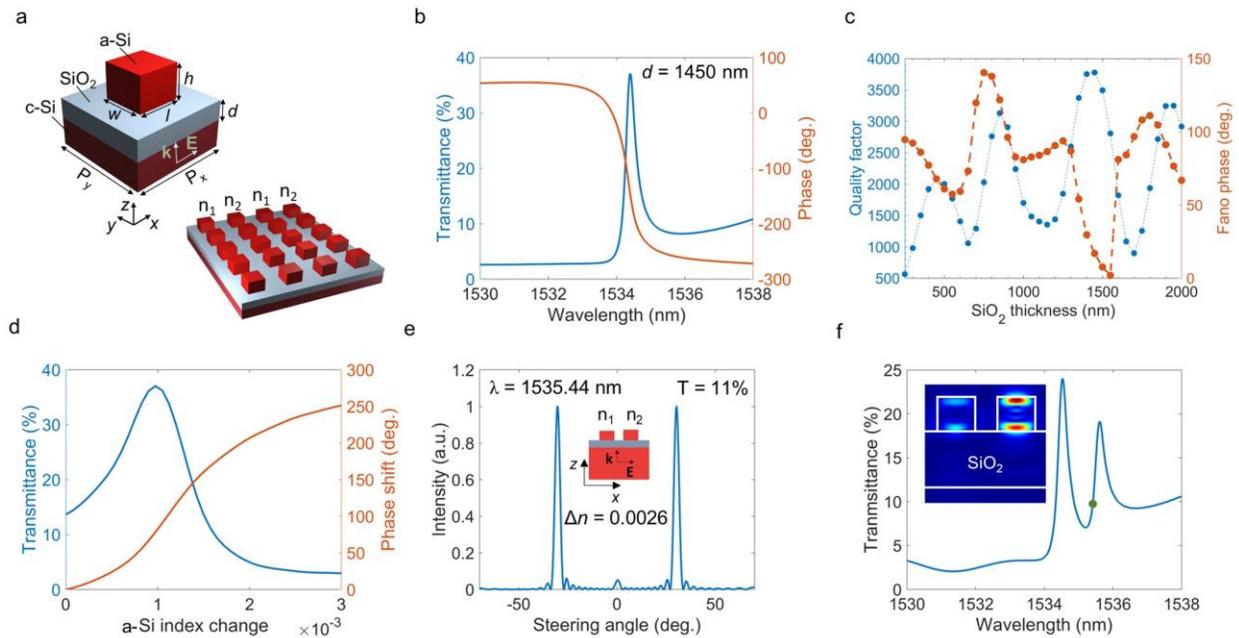

**Fig. 3**. a) Schematic of a transmissive metasurface. In a) the top panel shows the schematic of the metasurface unit cell while the bottom panel shows the metasurface. The proposed alternative design consists of a c-Si substrate, an SiO$_2$ spacer, followed by an a-Si rectangular pillar. b) Transmittance and phase spectra of the metasurface with a unit cell shown in a). In b), we assume that the thickness of the SiO$_2$ spacer $d$ is $d$ = 1450 nm. The height, width, and length of the a-Si pillar are $h$ = 845 nm, and $w$ = $l$ = 963 nm. The assumed period values are $P_x$ = 1520 nm and $P_y$ = 1425 nm. c) The resonance quality factor and Fano phase as a function of the SiO$_2$ thickness. d) Transmittance and phase shift as a function of the a-Si index change at a wavelength of $\lambda$ = 1534.8 nm. e) Intensity of the electric field in the far field for a two-level phase grating with 10 grating periods when the index change between neighboring a-Si pillars is $\Delta n$ = 0.0026. f) Overall transmittance of the two-level grating. In f), the green dot marks the transmittance at an operating wavelength of $\lambda$ = 1535.44 nm. The inset of f) shows the spatial distribution of the electric field in the x-z cross-section of the unit cell. The considered x-z cross-section goes through the center of the a-Si pillars.

The electric field profiles for a modified unit cell (Fig. 3a) exhibit mode leakage into the SiO$_2$ spacer (Supporting Information, Part 6). Both $E_x$ and $E_z$ components of the electric field have nonzero values in the SiO$_2$ spacer layer, and the electric field enhancement in a-Si pillars can be as high as 120. In addition to seeing changes in transmission, when changing the thickness of the SiO$_2$ spacer $d$, we observe variations in both the mode quality factor and Fano phase (Fig. 3c). By appropriately choosing the SiO$_2$ thickness $d$, we can tune the Fano phase over the range 0° to 140° (Fano phase is uniquely defined between 0° to 180°). This large variation of the Fano phase is indicative of comprehensive control of the resonant lineshape, exhibiting transmission dips, transmission peaks and asymmetric spectral shapes. Thus, by tailoring resonator mode leakage into the spacer layer, we modify the relative phase between the high-Q mode and the continuum, resulting in a change of the transmittance spectral lineshape.

As a next step, we assess metasurface beam steering (Fig. 3a). First, we consider the case in which all pillar refractive indices change by an equal amount Δ$n$ and calculate the phase shift and the transmittance as a function of the index change Δ$n$ at a given wavelength. Note the phase shift is defined as a difference between the phase of the transmitted light at Δ$n$ ≠ 0 and at Δ$n$ = 0. Our calculations show that a modest refractive index change of Δ$n$ = 3 × 10$^{-3}$ results in a phase shift of ~250° (Fig. 3d), which is sufficient to enable beam steering.

We consider the case of one-dimensional beam steering where the phases of all metasurface elements along the $y$ direction are identical while the phases of neighboring elements in the $x$ direction differ. Calculations for periodic arrays of identical elements suggest that an index change that would yield a 180° phase shift between neighboring elements in the $x$ direction can create a two-level phase grating and strongly suppress the zeroth order beam. However, our full wave simulations indicate that the index change value derived via this method is not optimal, as we obtain an unexpectedly large zeroth order beam. The observed low diffraction efficiency is due to non-negligible near-field coupling between neighboring metasurface elements. Our calculations indicate that an index change Δ$n$ = 0.0026 between neighboring rows of a-Si resonators at fixed operating wavelength of $\lambda$ = 1535.44 nm can fully extinguish the zeroth order diffraction beam (Fig. 3e). Interestingly, for these parameters ($\lambda$ = 1535.44 nm, Δ$n$ = 0.0026), the predicted relative phase shift between light scattered by adjacent rows of pillars is only 107°, according to the simulations of periodic arrays of identical resonators. In principle, this phase should result in a non-negligible zeroth order beam (Supporting Information, Part 6). These results indicate that such a simple ansatz for optimal beam steering does not readily apply for these high-Q metasurfaces.

Metasurface optical efficiency is a key design parameter; we find that overall transmittance of the two-level phase grating is T = 11% (Figs. 3e and 3f) which is significantly lower than the peak transmittance of ~35% derived from the identical resonator array simulations (Fig. S19). To understand the reduced transmittance, we assess the variation of two-level phase grating transmittance with wavelength (Fig. 3f).

The transmittance exhibits two distinct peaks indicating that the resonant wavelengths of neighboring metasurface unit cells have been shifted relative to each other by the imposed index change. The near-field distribution of the electric field (see the inset of Fig. 3f) also indicates that at a given operating wavelength, the relative electric field enhancement in neighboring a-Si rectangular pillars differs quite significantly. The optimal operating wavelength, which is marked by a green dot in Fig. 3f, is located between the two resonant peak wavelengths resulting in a lower overall transmittance.

The design of high-Q dynamic beam steering metasurfaces has two major challenges: i) One cannot use intuitive phase profile to steer the beam with a high steering efficiency, and the optimal beam steering conditions have to be identified via full wave simulations; ii) diffractive switching occurs only in one direction due to very significant inter-emitter near-field coupling in the direction perpendicular to the polarization of the incoming electric field (*y*-direction). Hence, it would be desirable to identify optical modes which enable two-dimensional beam steering[36].

Besides the high-Q mode, several lower-Q modes are supported with quality factors of ~200 (Supporting Information, Parts 3 and 4). The metasurface (Fig. 1) with a-Si rectangular pillar height, width, and length of $h$ = 845 nm, and $w$ = $l$ = 963 nm, respectively, and the period of $P_x$ = $P_y$ = 1500 nm, exhibits a relatively low-Q photonic mode at a wavelength of 1583 nm (Fig. 4a). As seen in Fig. 4a, the resonant dip observed at a wavelength of 1583 nm is also accompanied by a large spectral variation of the phase. When changing the refractive index of a-Si pillars by $\Delta n$ = 0.01 everywhere in the metasurface, we observe a dynamically tunable phase shift of ~285° (Fig. 4b), which is large enough to enable dynamical beam steering.

Next, we investigate whether the lower-Q mode can be used for dynamic beam steering, in two configurations: i) when the beam is steered in the direction perpendicular to the polarization of the incoming electric field (transverse electric or TE case); ii) when the steering direction is aligned with the polarization of the incoming light (transverse magnetic or TM case). First, we consider the case of a two-level phase grating and assume that the a-Si index change between neighboring element rows is $\Delta n$ = 0.0048 (for schematic of a grating period see insets of Figs. 4c and 4d). At $\Delta n$ = 0.0048, neighboring metasurface rows are expected to exhibit a phase difference of 180° at a wavelength of $\lambda$ = 1583.85 nm (Fig. 4b). Thus, we expect a near complete suppression of the specularly transmitted beam. However, our full wave simulations show significant transmission at normal incidence, for two reasons: i) near-field coupling between neighboring metasurface elements, ii) the difference in scattered light amplitude from neighboring elements[37].

To fully suppress the normally transmitted beam at an operating wavelength of $\lambda$ = 1583.85 nm, we developed an optimization procedure for the pillar refractive indices to maximize the diffraction efficiency of a desired order (see Methods), resulting in full suppression of the zeroth diffraction order at an operating wavelength of $\lambda$ = 1583.85 nm for both TE and TM polarizations of the incoming light (Figs. 4c and 4d). In the case of the TE polarization, the optimized refractive index of the one of the a-Si pillars within the grating period retained its original value ($n_1$ = 3.734), and the refractive index difference $\Delta n$ between neighboring a-Si pillars is $\Delta n$ = 0.0055, which yields a phase shift of 200° (see Fig. 4b), which is close to the originally chosen phase shift of 180°. In the case of the TM polarization, the optimization procedure yielded a-Si refractive indices: $n_1$ = 3.7353 and $n_2$ = 3.7440. The corresponding phase difference for adjacent a-Si pillars is 270°, which significantly deviates from the original value of 180°. For TM polarization, full suppression of the zeroth diffraction order occurs for a phase profile which significantly differs from the optimal phase shift uncoupled resonator array, indicating that inter-element near-field coupling is more significant for TM than TE polarization.

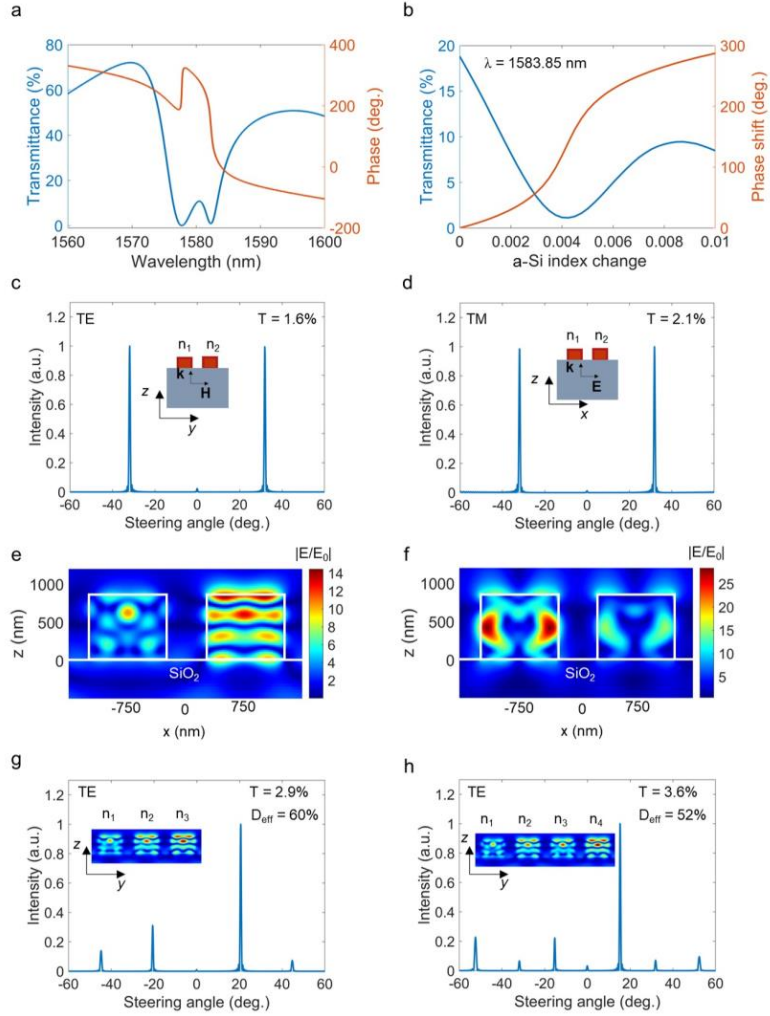

**Fig. 4.** Beam steering using the lower-Q mode. a) Transmittance and phase spectra of the metasurface with a unit cell shown in Fig. 1 b). In a), the assumed metasurface period values are $P_x$ = 1500 nm and $P_y$ = 1500 nm. The assumed geometrical parameters are as follows: the pillar height is $h$ = 845 nm, and the pillar length and width are $l$ = $w$ = 963 nm. b) Transmittance and phase shift as a function of the a-Si index change at a wavelength of $\lambda$ = 1583.85 nm. c) Intensity of the electric field in the far field as a function of the polar (steering) angle for a two-level phase grating when the incoming electric field is polarized perpendicular to the steering direction (TE polarization). d) Intensity of the electric field in the far field for a two-level phase grating when the incoming electric field is parallel to the steering direction (TM polarization). e) Near-filed electric field distribution in the two-level phase grating depicted in the inset of c), which corresponds to the case of the TE polarization. f) Near-filed electric field distribution in the two-level phase grating depicted in the inset d), which corresponds to the case of the TM polarization. In e) and f), $E_0$ denotes the amplitude of the impinging electric field. g) and h) plot the intensity of the electric field in the far field as a function of the steering angle in the cases of three-level and four-level phase gratings, respectively. g) and h) correspond to the case of the TE polarization. The insets of g) and h) show the near-field distribution of the electric field in each case. The overall transmittance values are specified in the insets of c), d), g), and h). The operating wavelength is $\lambda$ = 1583.85 nm.

Although we observed diffractive switching of a two-level phase grating, this analysis does not clarify whether intermediate steering angles are possible using a blazed grating-type design approach. In the case of TM polarization, using three- and four-level blazed grating phase profiles, we did not obtain a highly directional beam owing to near-field coupling between neighboring metasurface elements. However, for TE polarization, we demonstrated beam steering to angles of 20.6° and 15.2° with reasonable diffraction efficiencies. To steer a beam to a polar angle of 20.6°, we use Figure 4b to construct a phase a three-level phase profile with relative phase between elements of 0°, 120°, 240°, respectively, resulting in beam steering with a diffraction efficiency of 60% at an operating wavelength of $\lambda$ = 1583.85 nm (Fig. 4g). To steer the beam to a polar angle of 15.2°, a phase profile of 0°, 90°, 180°, 270° was constructed, resulting in a diffraction efficiency of 38% at an operating wavelength of $\lambda$ = 1583.85 nm and 46% at $\lambda$ = 1584.16 nm. To improve the diffraction efficiency at $\lambda$ = 1583.85 nm, we performed a full wave optimization procedure, yielding a diffraction efficiency of 52% (Figure 4h), with a-Si refractive indices: $n_1$ = 3.7340, $n_2$ = 3.7358, $n_3$ = 3.7386 $n_4$ = 3.7417. Thus, for TE-polarized incident light, our metasurface can steer the beam to intermediate angles between 0° and 20.6°.

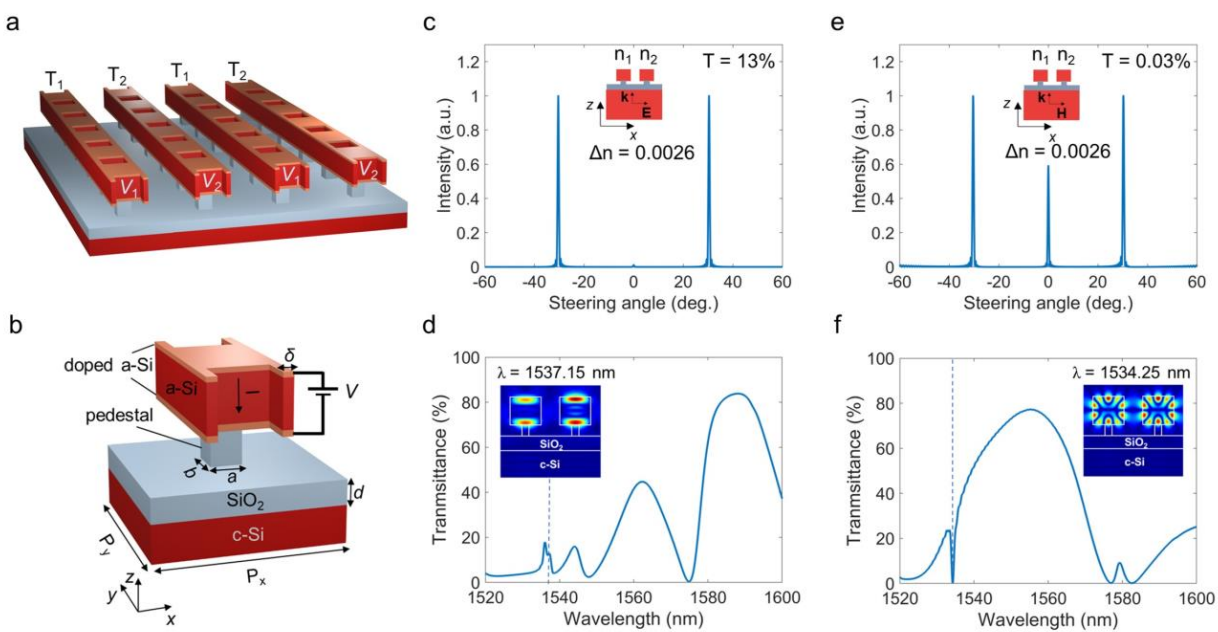

**Fig. 5.** Thermo-optic beam switching using the high-Q mode with a realistic interconnect architecture. In a), schematic of the proposed metasurface. In b), schematic of the proposed metasurface unit cell. In a)-b), the square a-Si pillars are connected via a-Si bars. Each a-Si pillar is placed upon an SiO$_2$ pedestal to enhance thermal insulation between neighboring metasurface elements. The width and length of the pillar is $w = l$ = 963 nm, the height of the pillar is $h$ = 850 nm. The width of the bar is $\delta$ = 50 nm, and the height of the bars is 850 nm. The SiO$_2$ pedestal has a shape of a rectangular pillar with a length and width $a = b$ = 200 nm, and a height of 382 nm. The thickness of the planar SiO$_2$ spacer is $d$ = 380 nm. The whole structure is built on an c-Si substrate. In c), the intensity of the electric field in the far field for a two-level phase grating when the incoming plane wave is *x* polarized. In c), the operating wavelength is $\lambda$ = 1537.15 nm. In d), the overall transmittance spectrum corresponding the two-level phase grating studied in c). In e), the intensity of the electric field in the far field for a two-level phase grating when the incoming plane wave is *y* polarized. In e), the operating wavelength is $\lambda$ = 1534.25 nm. In f), the overall transmittance spectrum corresponding the two-level phase grating studied in f). Insets of d) and f) show spatial distribution of the electric field in the *x-z* cross-section of the grating period at operating wavelengths in case of *x*- and *y*-polarized incoming light, respectively. The refractive index difference between neighboring metasurface elements is $\Delta n$ = 0.0026.

Having determined that both high- and lower-Q modes can be used for transmitted light wavefront manipulation, we seek a feasible design to enable refractive index variation and dynamic beam switching, by refractive index modulation of a-Si using the thermo-optic effect [14,23,38]. To selectively heat rows of a-Si pillars, we introduce a-Si electrodes and connect the pillars in series (Figs. 5a and 5b). In our design, we place a-Si pillars on $SiO_2$ pedestals to limit thermal crosstalk between neighboring pillars. Fabrication entails first patterning a-Si on an $SiO_2$ spacer followed by hydrofluoric acid wet etching of the $SiO_2$ spacer into pillars on a c-Si substrate.

Figures 5a and 5b illustrate this design - the top and bottom 50 nm-thick a-Si layers are doped, and a voltage is applied to the top and bottom layer resulting in current flow between the electrodes through the lightly doped a-Si layer (see Fig. 5b). The induced current raises the temperature $T$ of a-Si pillars by Joule heating, with modified a-Si refractive index $n(T) = n_{Si} + \Delta n(\Delta T)$. For silicon, a temperature change of $\Delta T$ = 10 K changes the refractive index by $\Delta n$ = 0.00239 [39]. Dynamic beam switching using a higher-Q mode is possible with refractive index difference as low as $\Delta n$ = 0.0026, corresponding to a relative temperature difference of ~11 K. For beam switching with lower-Q modes, a refractive index difference of $\Delta n$ = 0.006 is required, corresponding to a relative temperature difference of $\Delta T$ = 25 K. These numerical estimates suggest our metasurface can potentially enable beam steering with quite modest temperature differences.

To construct a desired transmitted light phase profile, we can heat individual rows of pillars to different temperatures. For the high-Q mode, diffractive switching with a two-level phase grating can be realized by changing the temperature of every other metasurface pixel (i.e., electrically connected rows of metasurface elements) by 11 K ($\Delta n$ = 0.0026). Heating to create a two-level phase grating, deflects the light to angles of $\theta$ = ±30° (Figs. 5c and 5e). For an incident $x$-polarized electric field, this modest refractive index difference enables diffractive beam switching with a complete suppression of the zero[th] diffraction order and an overall transmittance of $T$ = 13 % (Fig. 5c). We also observe diffractive beam switching for $y$-polarized incidence (Fig. 5d) with a lower overall transmittance and at a different operating wavelength. For a-Si on an $SiO_2$ substrate, we only observed diffractive switching for the high-Q mode when the steering direction aligned the incoming plane wave (TM) polarization. The a-Si electrodes and the $SiO_2$ pedestal serve to reduce near-field coupling of neighboring metasurface elements. Lower-Q modes are also capable of diffractive switching for both TE and TM polarized incidence, at different wavelengths.

We also performed a coupled electrical and thermal analysis (see Methods). To induce a diffraction grating analogous to the one in Fig. 5a, we apply a voltage of $V_1$ = 1.05 V between top and bottom doped a-Si layers of one metasurface pixel while grounding each neighboring metasurface pixel ($V_2$ = 0 V) in a periodic array. Figure 6a illustrates the calculated current density in the silicon bars in one grating period. The largest current density and most significant heat generation occurs in the top and bottom doped a-Si layers of the connector bars. Figure 6b illustrates the steady state temperature distribution for one grating period for $V_1$ = 1.05 V and $V_2$ = 0 V, such that the temperatures of two adjacent metasurface pixels are $T_1$ = 312.8 K and $T_2$ = 301.4 K, respectively, yielding refractive index difference of $\Delta n$ = 0.0026, assuming a thermo-optic coefficient of 2.3 × $10^{-4}$ $K^{-1}$ [39]. The power density required to maintain this temperature difference in steady state is 1.62 µW/µm². The temperature distribution in the pillars and connectors is nearly uniform due to the high thermal conductivity of silicon, even though heating occurs predominantly in the connector bars. The largest temperature gradient occurs in the silicon oxide pedestal due to its large thermal resistance, which greatly reduces the thermal crosstalk between adjacent rows. To assess metasurface dynamic performance, we carried out transient electrical and thermal simulations of dynamic thermal switching between the two pixels comprising one grating period. A response time of 7.3 µs is obtained for a square wave voltage profile (see Fig. 6c), indicating that thermo-optic switching is possible at frequencies up to 140 kHz.

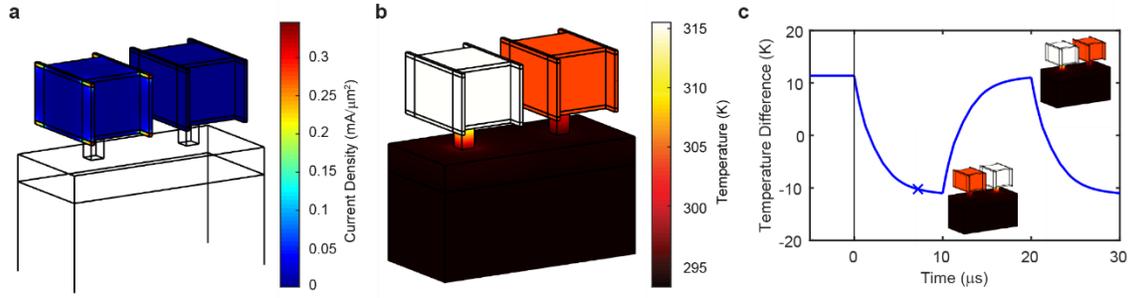

**Fig. 6.** Electrical and thermal analysis of thermo-optic beam switching. a) Steady state spatial current density distribution in a thermo-optically controlled metasurface with $V_1$ = 1.05 V and $V_2$ = 0 V. b) Steady state spatial temperature distribution in a thermo-optically controlled metasurface with $V_1$ = 1.05 V and $V_2$ = 0 V. c) Transient temperature difference between the two rows of a thermo-optically controlled metasurface when switching the voltage with a square wave of amplitude $V$ = 1.05 V and a frequency of 100 kHz. The geometrical parameters are identical to the ones of the structure in Fig. 5.

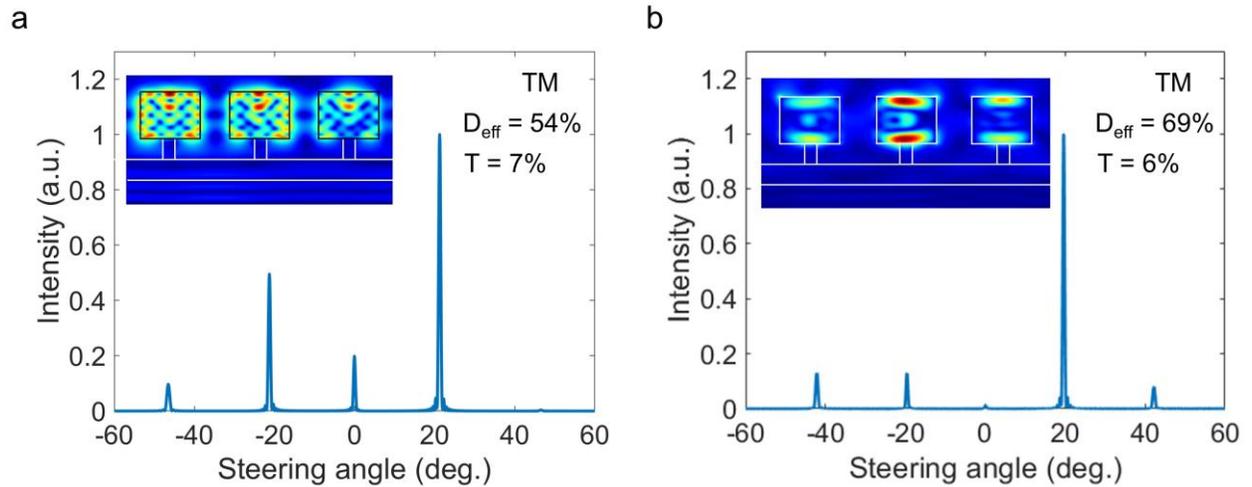

**Fig. 7.** Thermo-optic three-level phase grating with a realistic interconnect architecture and optimized geometry. The incoming plane wave is TM-polarized. a) Intensity of the electric field in the far field for the case of the lower-Q mode. In a), the operating wavelength is λ = 1567.1 nm. b) Intensity of the electric field in the far field for the case of the high-Q mode. In b), the operating wavelength is λ = 1531.3 nm. The insets of a) and b) show the spatial distribution of the electric field in the x-z cross-section of the grating period at the operating wavelengths.

Finally, we assess metasurface beam steering performance with realistic interconnects (Fig. 5a) when a three-level phase profile (0°, 120°, 240°) is applied to the metasurface, for both lower-Q and high-Q modes under TM-polarized plane wave illumination. This phase profile yields a maximum diffraction efficiency of 35% for the lower-Q mode and 28% for the high-Q mode (see Supporting Information). Optimization (see Methods) of the a-Si pillar refractive indices resulted in a modest improvement of diffraction efficiencies. To further increase diffraction efficiency, we co-optimized the geometrical parameters of the structure and refractive indices of a-Si pillars (see Methods) enabling enhanced diffraction efficiencies of 54% and 69% for the lower-Q and the high-Q mode, respectively (see Fig. 7).

**Conclusions**

In summary**,** all-dielectric dynamically tunable transmissive metasurfaces are achievable using high-Q resonances with Q-factors approaching 10,000. Appropriate design enables the metasurface to exhibit a transmittance *peak* (rather than a transmittance dip at the resonant wavelength for Q-factors of ~3780 (Fig. 3a). These high quality factors enable achieving beam steering at modest variations of the a-Si refractive index on the order of $10^{-3}$, which is a unique advantage of the proposed design. Lower-Q modes can also be used for amplitude and phase modulation, but the required a-Si refractive index modulation is higher (up to 0.01) as compared with the case of high-Q mode. Thermo-optically active elements can be addressed using interconnects that do not perturb the resonator cavity modes, and both high-Q and lower-Q modes can realize thermo-optic beam switching, with switching times as low as 7.3 µs. For both high-Q and lower-Q modes, dynamic switching of both TE- and TM-polarized light is possible, although at different wavelengths. Using a three-level phase grating approach, we have shown that the metasurface with realistic interconnects is capable of dynamic beam steering, with diffraction efficiencies of 54% and 69% for lower Q and high-Q modes, respectively. Finally, we note that the optical efficiency of the proposed metasurface can be significantly enhanced by introducing a metallic back reflector and an $SiO_2$ spacer. The designed reflective metasurfaces may exhibit optical efficiencies >90% (Supporting Information, Part 9).

**Methods**

Optical simulations were performed using the finite difference time domain method (FDTD Lumerical). In our optical simulations, a normally incident linearly polarized plane wave illuminates the metasurface from within the substrate (see Figs. 1 and 2). When simulating an array of a-Si pillars, periodic boundary conditions are used in the *x* and *y* directions, and perfectly matched layers (PML) boundary condition is used in the *z* direction. When considering the behavior of an isolated pillar, PML boundary conditions are used at all simulation boundaries, and the simulation volume is $3 \times 3 \times 3$ µm$^3$. In our simulations the materials are assumed non-dispersive. The refractive indices of a-Si, $SiO_2$, and c-Si are taken as 3.734 [35], 1.44, 3.43, respectively. When considering realistic interconnect architectures to perform thermo-optic beam switching and thermo-optic beam steering, we assume the complex refractive index of the doped a-Si layers is given as $n = 3.734+0.0013 i$, which corresponds to the carrier density of $6 \times 10^{18}$ cm$^{-3}$ in the case of the *n*-doped Si layer and the carrier density of $10^{19}$ cm$^{-3}$ in case of the *p*-doped Si layer [40]. Complex refractive index of the lightly doped Si core is $n = 3.734+0.000025 i$, which corresponds to the carrier density of $3.2 \times 10^{17}$ cm$^{-3}$ of the *n*-doped Si [40]. In simulations used to generate Figs. 1 and 2, the assumed mesh in the *z*-direction is 5 nm while the mesh in the *x*- and *y*-directions is set to 20 nm. In our beam switching and beam steering simulations, the mesh in the *x*-, *y*-, and *z*-directions is set to 20 nm. When performing the near to far field projection in Fig. 3, we assume that the number of metasurface elements in the *x*-direction is 20, while for the near to far field projection in Figs. 4, 5, and 7, the number of assumed metasurface elements in the x-direction is 100.

To optimize diffraction efficiencies, we used MATLAB to drive a multi-variable nonlinear optimization in FDTD via Lumerical's Automation Application Programming Interface (API). When performing the optimization, we used the sequential quadratic programming (SQP) method and used the inverse of the maximal diffraction efficiency as a figure of merit. The phase profile was assumed periodic, and the geometry and the refractive indices of the structure within a period were varied. When optimizing diffraction efficiencies in Figure 4, only refractive indices of a-Si pillars were varied so that the maximal assumed index change was $\Delta n = 0.01$. When optimizing diffraction efficiencies of the structure with realistic interconnects (Fig. 7) we run a series of optimization runs. First, we co-optimized only refractive indices of the metasurface elements. In the next step, we fixed the refractive index of the metasurface element $n_1$ and co-optimized the structure height and the refractive indices of the two remaining metasurface elements. Next, we co-optimized the $P_y$ period (or $P_x$ period) of the structure and refractive indices of the second and third metasurface element while keeping the refractive index of the first metasurface element fixed. As a final optimization step, we re-optimized the refractive indices of all three metasurface elements. In the case of the lower-Q mode, the optimization yields the following values for the refractive indices of the Si pillars and the geometrical parameters of the structure: $n_1 = 3.73416$, $n_2 = 3.738299$, $n_3 = 3.73778$, $P_x = 1440$ nm, $P_y = 1440$ nm, and $h = 840.75$ nm, and the observed diffraction efficiency $D_{eff}=54\%$ (Fig. 7a). In the case of the high-Q mode, the obtained parameter values are as follows: $n_1 = 3.73416$, $n_2 = 3.7383$, $n_3 = 3.744$, $P_x = 1520$ nm, $P_y = 1495.92$ nm, and $h = 841.207$ nm, and the resulting diffraction efficiency $D_{eff} = 69\%$ (Fig. 7b).

Three-dimensional electrical and thermal simulations were performed using finite element method (COMSOL Multiphysics). In our electrical simulations, we assume that, in the metasurface unit cell, the top and bottom 50 nm of a-Si are *n*-doped with assumed carrier density of $6 \times 10^{18}$ cm$^{-3}$. The carrier density of the lightly doped core is taken to be $3.2 \times 10^{17}$ cm$^{-3}$. First, we use electrical solver to obtain the volumetric heat source distribution due to Joule heating. As a next step, we use a thermal solver, to obtain the temperature distribution. Our thermal simulations account for heat conduction and convection. We assume that the top of the metasurface is cooled via natural convection with a heat transfer coefficient of $h = 5$ W/m$^2$K. We also assume that the temperature at the bottom of the substrate, 50 mm from the metasurface is fixed to 298 K with an external heat sink. The assumed ambient temperature is also 298 K. When performing thermal simulations, we use periodic boundary conditions in the *x* and *y* directions.

**Author information**

**Corresponding Author:**

Harry A. Atwater

*E-mail: haa@caltech.edu

**Acknowledgements**

This work was supported by grant #FA9550-18-1-0354 from Air Force Office of Scientific Research as well as the Meta-Imaging MURI grant #FA9550-21-1-0312 from Air Force Office of Scientific Research.

**Author contributions**

R.S., H.A.A., and C.U.H. conceived the project. R.S. designed the metasurface and preformed optical simulations. M.F. extracted Q-factors and Fano phase from the simulated transmittance spectra and performed a finite array analysis. C.U.H. performed electrical and thermal simulations and devised strategies minimizing thermal crosstalk. R.S. and M.Y.G. performed the full wave optimization. C.U.H., M.Y.G., and M.F contributed to the discussion of the results. R.S. wrote the manuscript with inputs from other authors. H.A.A supervised all aspects of the project.

**Competing interests**

The authors declare no competing interests.

**Data Availability Statement**

The data that support the findings of this study are available from the corresponding author upon reasonable request.

# Supporting Information

1. **Extraction of the quality factor and Fano phase**

We extracted the quality factor and Fano phase values by fitting the transmittance spectra to the Fano formula [1]:

$$T = A \frac{\left|\frac{\gamma_0}{2} - e^{-2i\Delta}\left(i(\omega - \omega_0) - \frac{\gamma_0}{2}\right)\right|^2}{(\omega - \omega_0)^2 + \left(\frac{\gamma_0}{2}\right)^2} + T_{bg}$$

Here, $T_{bg}$ is a constant offset plus a linear background ($T_{bg} = B + C(\omega - \omega_0)$), $A$ is the resonance amplitude, $\omega$ denotes the frequency of light while $\omega_0$ is the resonant frequency, $\gamma_0$ is the damping constant. $\Delta$ denotes the Fano phase, and the Fano asymmetry parameter $q$ is related to the Fano phase Δ as $q = -\cot(\Delta)$ [2]. The quality factors are calculated as $Q = \frac{\omega_0}{\gamma_0}$.

## 2. Field profile of the high-Q mode in the pillar array

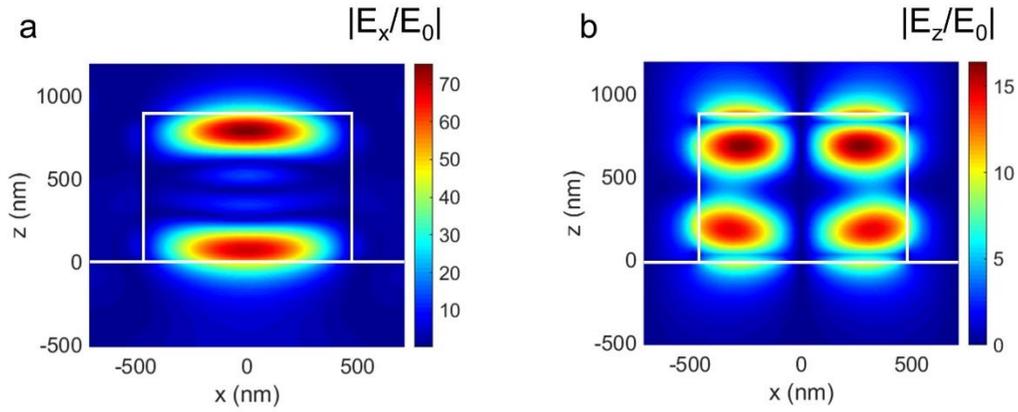

Fig. S1. Spatial distribution of the electric field amplitude inside the metasurface unit cell shown in Figure 1, which described the case of an array of a-Si pillars on an SiO$_2$ substrate. Namely, the length and width of the pillar are taken to be $l = w$ = 963 nm. The metasurface period is $P_x = P_y$ = 1425 nm. The electric field is plotted in the $x$-$z$ plane, which goes through the center of the pillar (the same as in Fig. 1d). a) $x$-component of the electric field, b) $z$-component of the electric field. Field profiles are plotted at a resonant wavelength of the resonance dip shown in Fig. 1c. The $E_y$ component is identically equal to zero ($E_y$ = 0).

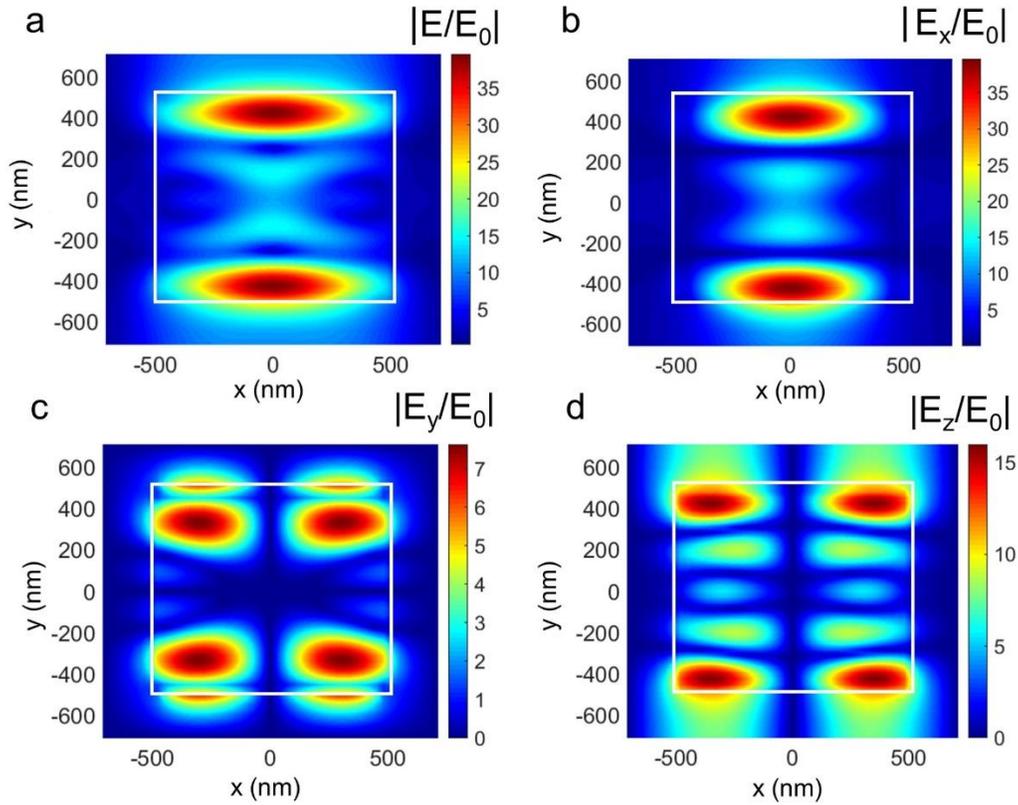

Fig. S2. Spatial distribution of the electric field amplitude inside the metasurface unit cell shown in Figure 1., which described the case of an array of a-Si pillars on an SiO$_2$ substrate. The length and width of the pillar are taken to be $l = w$ = 963 nm. The metasurface period is $P_x = P_y$ = 1425 nm. The electric field is plotted in the x-y plane, which goes through the center of the pillar. a) An absolute value of the electric field, b) x-component of the electric field, c) y-component of the electric field, d) z-component of the electric field.

## 3. Optical response of a pillar array: effect of the pillar height on the array performance

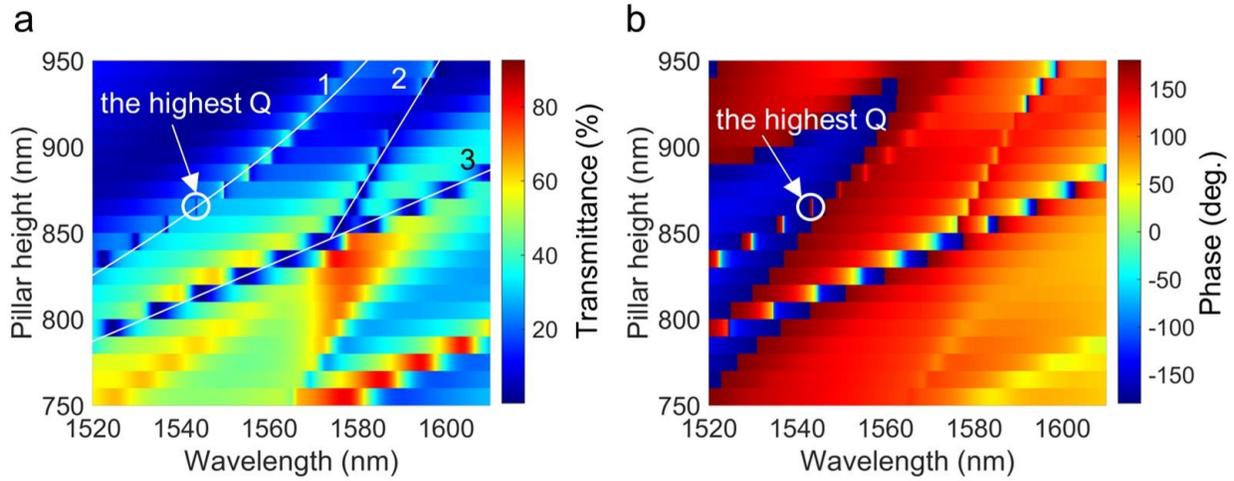

Fig. S3. Optical response of a-Si pillar array on an $SiO_2$ substrate. The metasurface design is the same as in Figs. 1 and 2. Namely, the length and width of the pillar are taken to be $l = w = 963$ nm. The metasurface period is $P_x = P_y = 1425$ nm. Transmittance a) and the phase of a transmitted light b) as a function of wavelength and the a-Si pillar height $h$. We observe that modes 2 and 3 coalesce at a pillar height of $h = 850$ nm, and the highest-Q high-Q mode is observed at a pillar height of $h = 860$ nm.

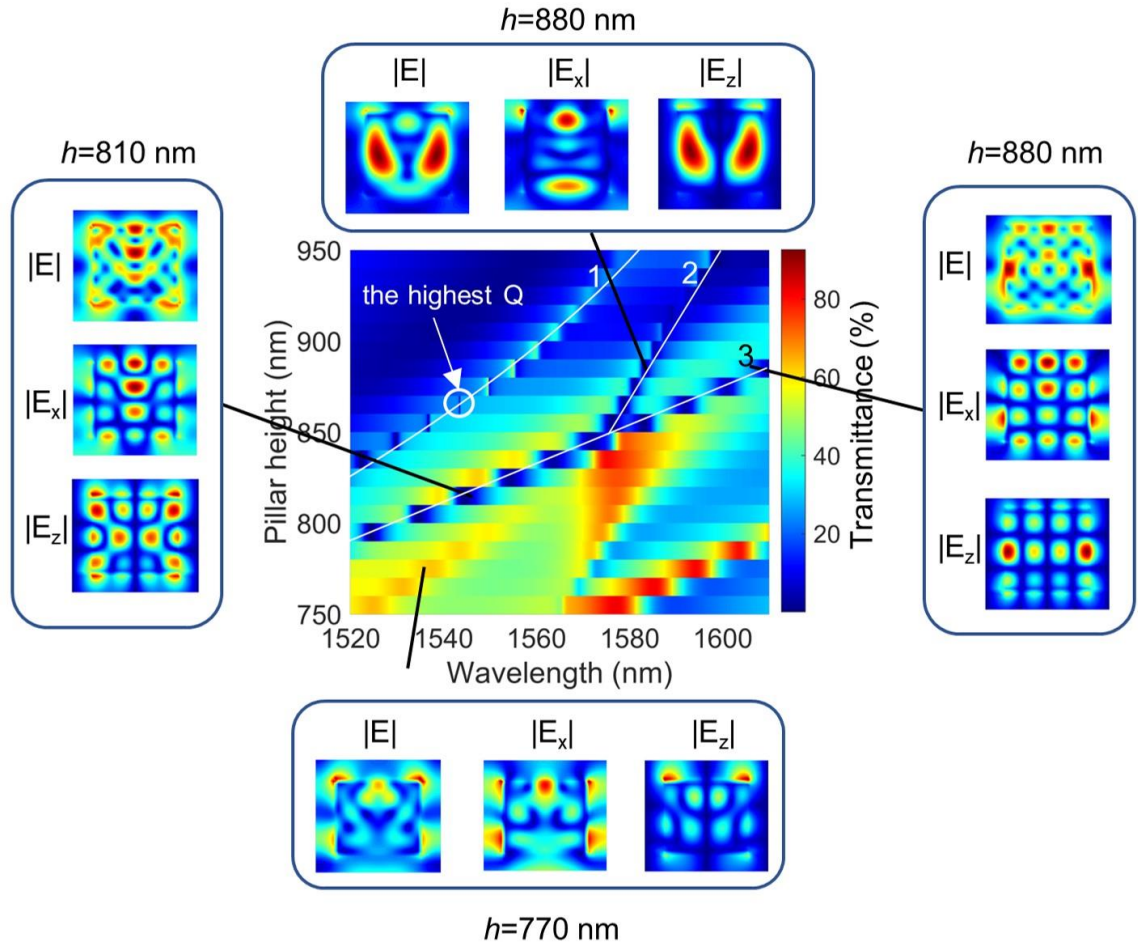

Fig. S4. Spatial electric field profiles of a lower Q modes supported by the metasurface in the ***x-z* plane**, which passes through the center of the pillar. In the displayed electric field profiles, the *z* coordinate ranges from *z* = -200 nm to *z* = 1000 nm while the top of the $SiO_2$ substrate corresponds to the plane *z* = 0. The *x* coordinate ranges from *x* = -712.5 to *x* = 712.5 nm. The displayed false color plot is identical to Fig. S3a. The spatial field profile of the high-Q mode is identical to the one shown in Figs. 1, S1, and S2. We observe that *x*- and *y*-components of the electric field of mode 3 have very similar features when varying the pillar height. On the other hand, mode 2 'disappears' after coalescing with mode 3. The mode profile of mode 1 is identical to the one shown in Figs. 1, and S1.

-

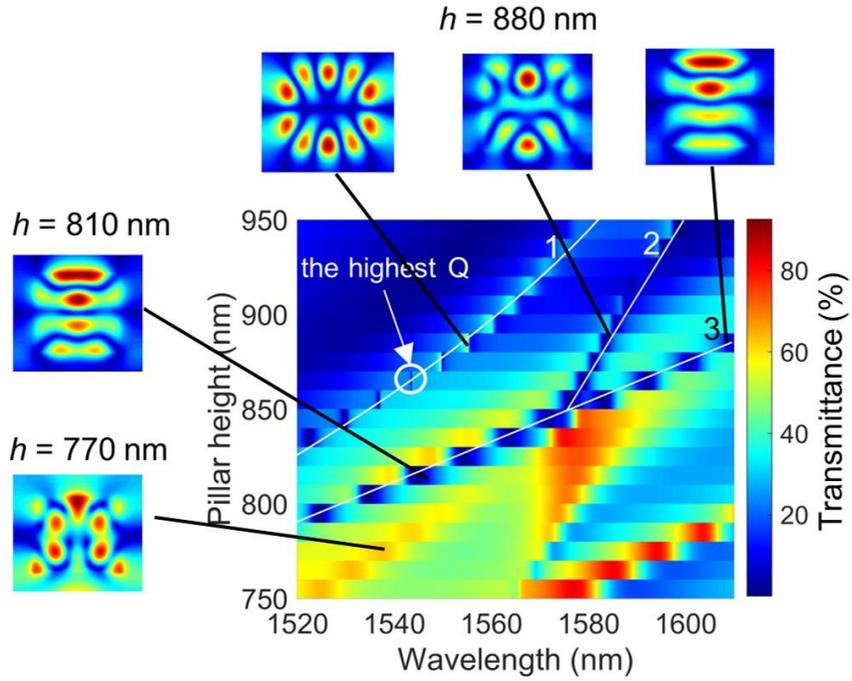

Fig. S5. Spatial electric field profiles of a lower Q modes supported by the metasurface in the **y-z plane**, which passes through the center of the pillar. In the displayed electric field profiles, the $z$ coordinate ranges from $z = -200$ nm to $z = 1000$ nm while the top of the $SiO_2$ substrate corresponds to the plane $z = 0$. The $x$ coordinate ranges from $x = -712.5$ nm to $x = 712.5$ nm. The displayed false color plot is identical to Fig. S3a. The spatial field profile of the high-Q mode is identical to the one shown in Figs. 1, S1, and S2. We observe that $x$- and $y$-components of the electric field of mode 3 have very similar features when varying the pillar height. On the other hand, mode 2 'disappears' after coalescing with mode 3. The mode profile of mode 1 is identical to the one shown in Figs. 1, and S1. In this Figure, we display electric field amplitude $|E|$. On the considered $y$-$z$ plane, the only non-zero component is $E_x$.

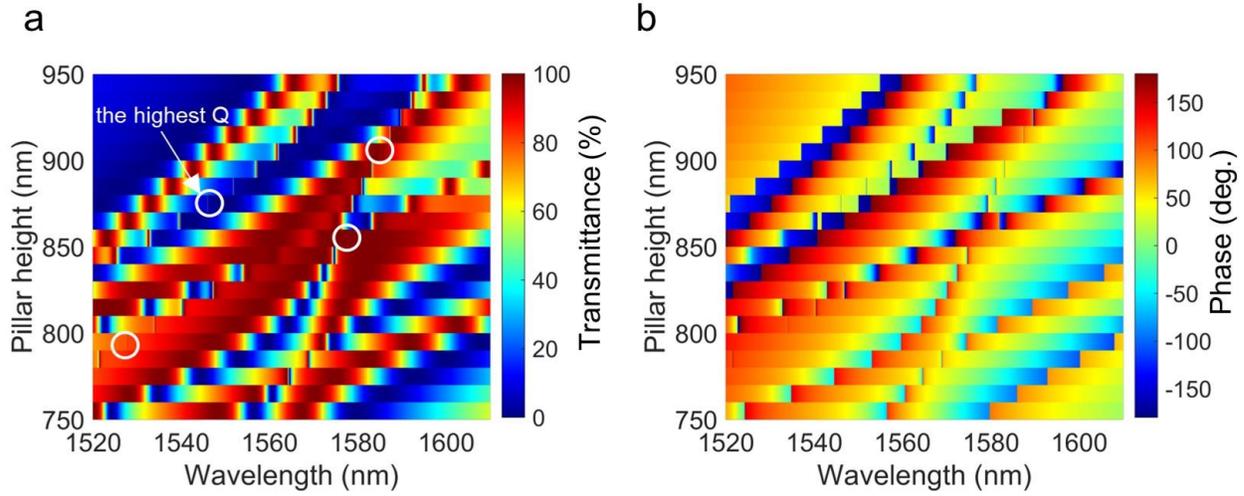

Fig. S6. Optical response of a-Si pillar array suspended in air. The assumed geometrical parameters are the same as in Fig. 1. Namely, the length and width of the pillar are taken to be $l = w = 963$ nm. The metasurface period is $P_x = P_y = 1425$ nm. In a), the transmittance as a function of wavelength and the a-Si pillar height $h$. In b), the phase of the transmitted light as a function of wavelength and the a-Si pillar height $h$. Compared with the case of pillars on an $SiO_2$ substrate, we observe the abundance of extremely high-Q modes, which are marked by a white circle in a). At the pillar height of $h = 870$ nm, the Q-factor of the observed high-Q mode is ~48,000.

When considering an a-Si pillar array we observe that at a pillar height of 870 nm, we are still able to couple to the resonant mode, and the resonant spectral feature is still visible from the false color plots of the transmittance and phase spectra (Fig. S6). The extracted Q-factor of the resonance is ~48,000 (Fig. S7a). In our simulation, we assumed a 5 nm mesh in z direction. Running a series of simulations with finer mesh could potentially enable us to identify the parameter values at which the mode with even higher Q-factor can be observed. We also observe that when take the metasurface period as $P_x = 1520$ nm and $P_y = 1425$ nm, the Q-factor of the supported mode is 221,000 (Fig. S7b). Thus, increasing the metasurface period from $P_x = 1425$ nm to $P_x = 1520$ nm can strongly affect the quality factor of the metasurface.

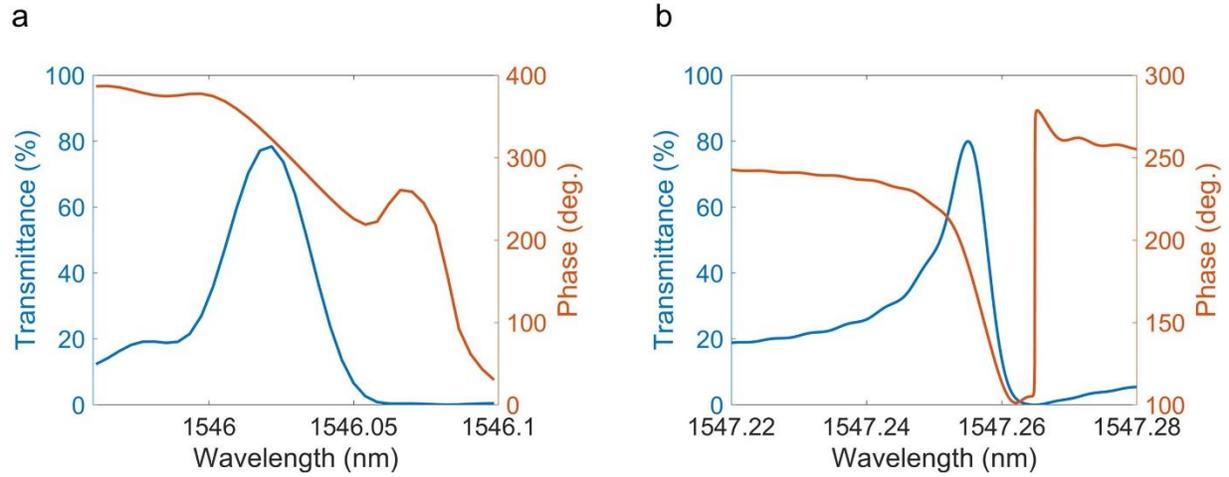

Fig. S7. Transmittance and phase spectrum for an array of a-Si pillars *in air*. The assumed width, length and height of the pillars are $w = l$ = 963 nm, $h$ = 870 nm (see the schematic in Fig. 1). In a), the values for the period are $P_x$ = 1425 nm and $P_y$ = 1425 nm. The extracted Q-factor of the resonance is ~48,000. The mesh is 20 nm in the *x*- and *y*-directions, and 5 nm in the *z*-direction. In b), the values for the period are as compared with a). In b), $P_x$ = 1520 nm and $P_y$ = 1425 nm. To reduce simulation time, we assumed a mesh of 20 nm in all three directions. The extracted Q-factor of the resonance is ~211,000.

## 4. Optical response of a pillar array: effect of the pillar period on the array performance

We study how the period of the metasurface affects the modes supported by the metasurface consisting of an array of a-Si pillars on an $SiO_2$ substrate (Fig. 1). For metasurface periods exceeding 1200 nm, we observe three distinct modes in the transmittance and phase false color plots, which correspond to the high-Q mode at wavelength around 1540 nm and two lower-Q modes at wavelengths around 1580 nm and 1590 nm, which correspond to mode 2 and mode 3 from Fig. S3. When studying the dependence of the mode position on the x-period $P_x$, we observe that the quality of the high-Q mode gradually increases with period (Fig. S8c and S9c). We also observe that for periods of $P_x > 1300$ nm, the position of the high-Q mode is does not change significantly with period. We also observe that mode 2 shifts stronger with the x-period $P_x$ as compared with mode 3. We also observe, that when we change the y-period $P_y$, the high-Q mode shifts stronger as compared with the case when the y-period is changed (c.f. Figs. S8c and S9c). This result is also consistent with the specifics of the spatial mode profiles of the high-Q mode in the *x-z* and *y-z* planes (see Fig. 1). When the *y*-period $P_y$ increases from 1300 nm to 1520 nm, the high-Q resonance position shifts by ~5 nm.

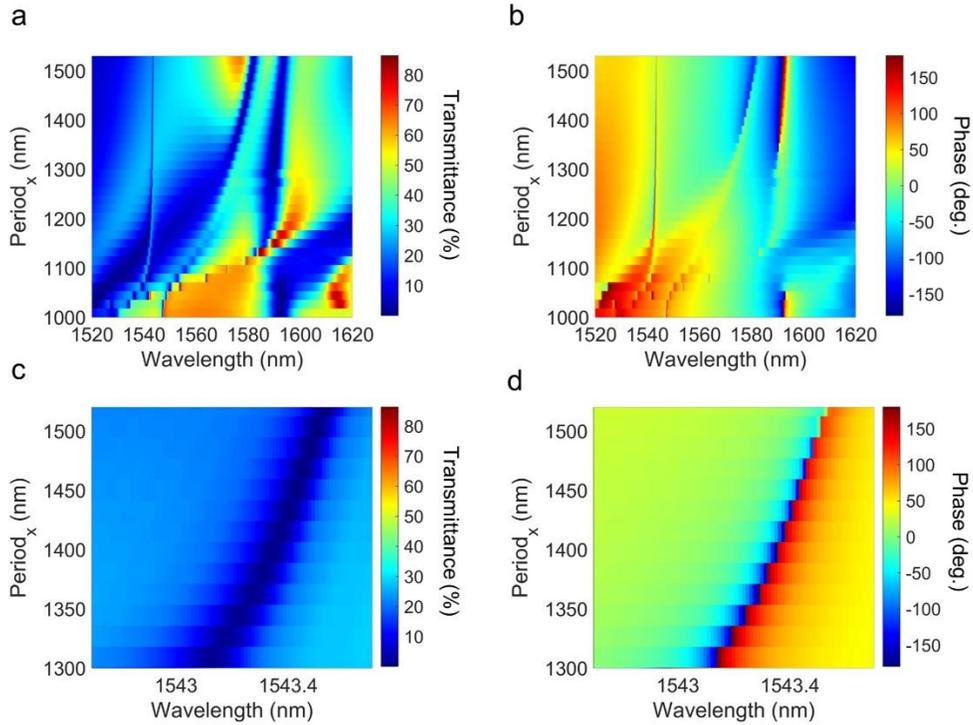

Fig. S8. Optical response of an a-Si pillar array on an $SiO_2$ substrate when the incoming electric field is x-polarized (see Fig. 1). The metasurface design is the same as in Figs. 1 and 2. Namely, the length and width of the pillar are taken to be $l = w = 963$ nm, the height of the pillar is $h = 860$ nm. The metasurface period in the y-direction is $P_y = 1425$ nm. Transmittance a) and the phase of a transmitted light b) as a function of wavelength and the period in the x-direction $P_x$. c) and d) take a closer look at the behavior of the high-Q mode by limiting the range of the x-period to [1300 nm, 1520 nm] in a) and b), respectively. In c) and d), the wavelength range is limited to [1542.7 nm, 1543.7 nm].

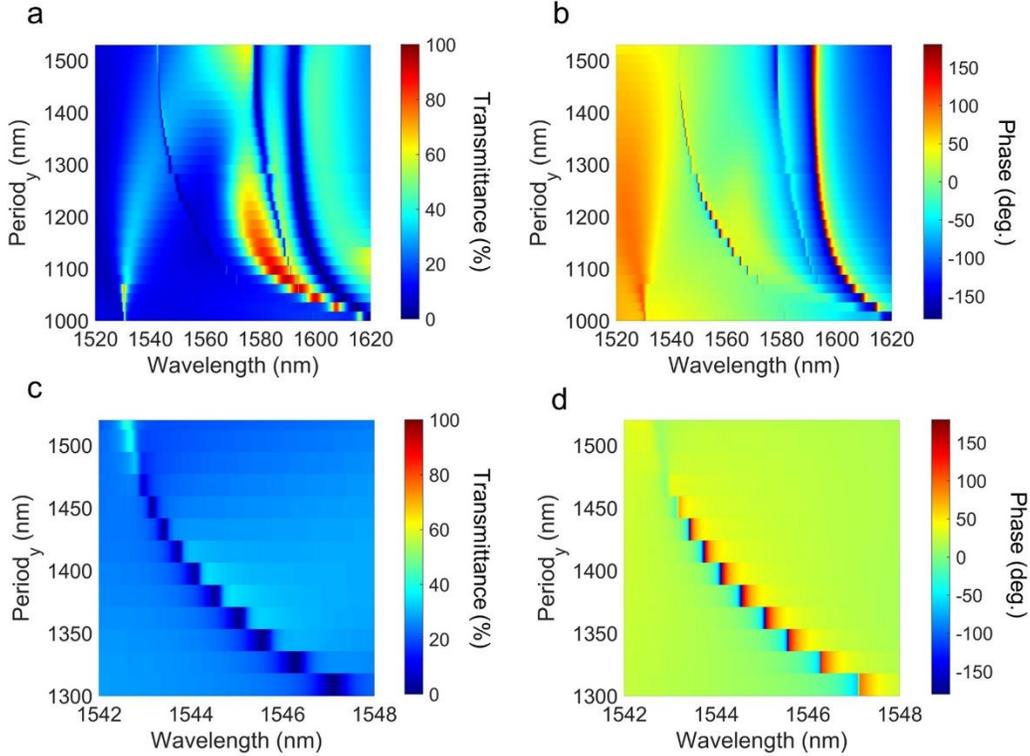

Fig. S9. Optical response of an a-Si pillar array on an $SiO_2$ substrate when the incoming electric field is *x*-polarized (see Fig. 1). The metasurface design is the same as in Figs. 1 and 2. Namely, the length and width of the pillar are taken to be *l* = *w* = 963 nm, the height of the pillar is *h* = 860 nm. The metasurface period in the *x*-direction is $P_x$ = 1425 nm. Transmittance a) and the phase of a transmitted light b) as a function of wavelength and the period in the *y*-direction $P_y$. c) and d) take a closer look at the behavior of the high-Q mode by limiting the range of the *y*-period to [1300 nm, 1520 nm] in a) and b), respectively. In c) and d), the wavelength range is limited to [1542 nm, 1548 nm].

## 5. Modes supported by a single isolated pillar

So far, in our optical simulations we imposed periodic boundaries in x- and y- directions implying that we investigate a periodic array of square a-Si pillars. It is not clear whether the identified photonic mode is also supported by an isolated a-Si pillar. To address this question, we simulate scattering cross section of an isolated pillar. In the revised simulation, we use perfectly matched layer (PML) boundary conditions at all simulation boundaries. First, we consider the case of a *single a-Si pillar in air* and calculate its scattering cross section as a function of wavelength and pillar height (Fig. S10a). The length and width of the pillar are the same as in Fig. 1 of the main manuscript ($l = w = 963$ nm). As seen in Fig. S10, in the vicinity of the highest-Q mode we also observe crossing of two other modes. For a pillar height of $h$ = 834 nm, the quality factor of the highest-Q mode is ~1000.

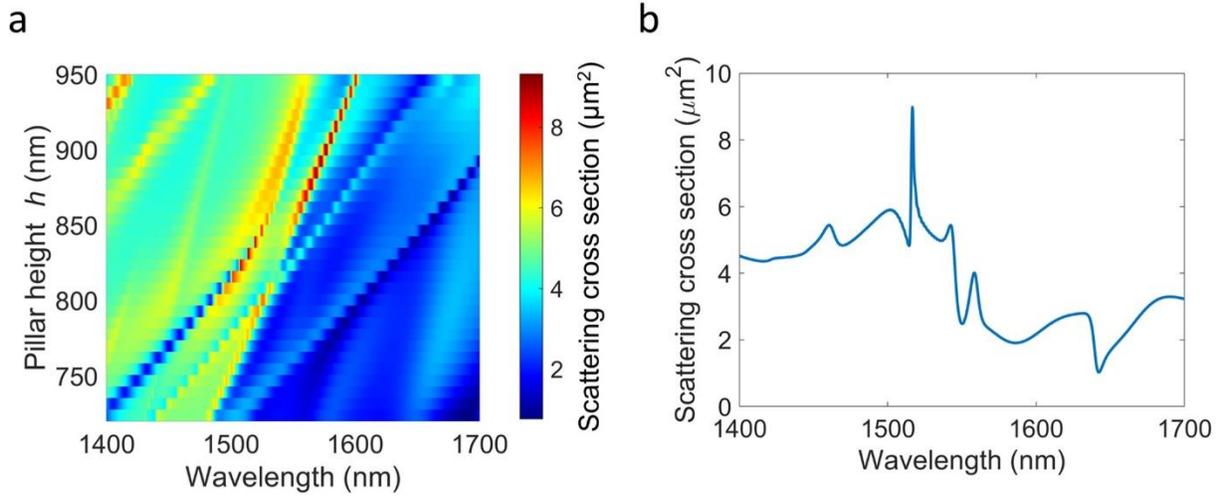

Fig. S10. a) Scattering cross section of a single a-Si pillar *in air* as a function of wavelength and pillar height. The length and width of the pillar is $l = w = 963$ nm. b) Scattering cross section of a single a-Si pillar in the air where the pillar height is $h = 834$ nm. The highest-Q mode is observed at a wavelength of $\lambda = 1519.8$ nm.

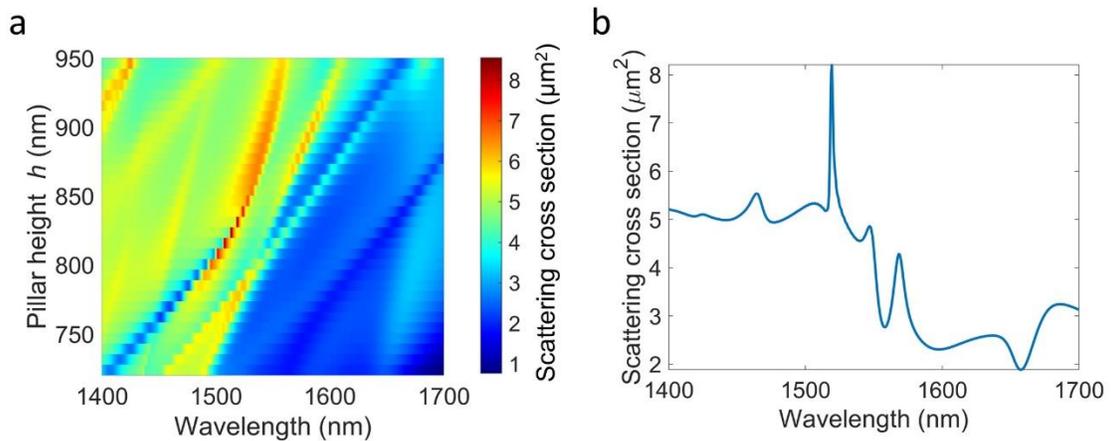

Fig. S11. a) Scattering cross section of a single a-Si pillar on an *SiO$_2$ substrate* as a function of wavelength and pillar height. The length and width of the pillar is $l = w = 963$ nm. b) Scattering cross section of a single a-Si pillar where the pillar height is $h = 830$ nm. The highest-Q mode is observed at a wavelength of $\lambda = 1519.2$ nm.

Next, we study the case of an isolated a-Si pillar on an $SiO_2$ substrate. Figure S11 plots the scattering cross section of an isolated pillar on an $SiO_2$ substrate as a function of wavelength and pillar height (Fig. S11a). We observe an overall broadening of the spectral features as compared with the case of a single a-Si pillar in the air. We also observe that for a pillar height of 830 nm, the Q-factor of the supported mode is 676.

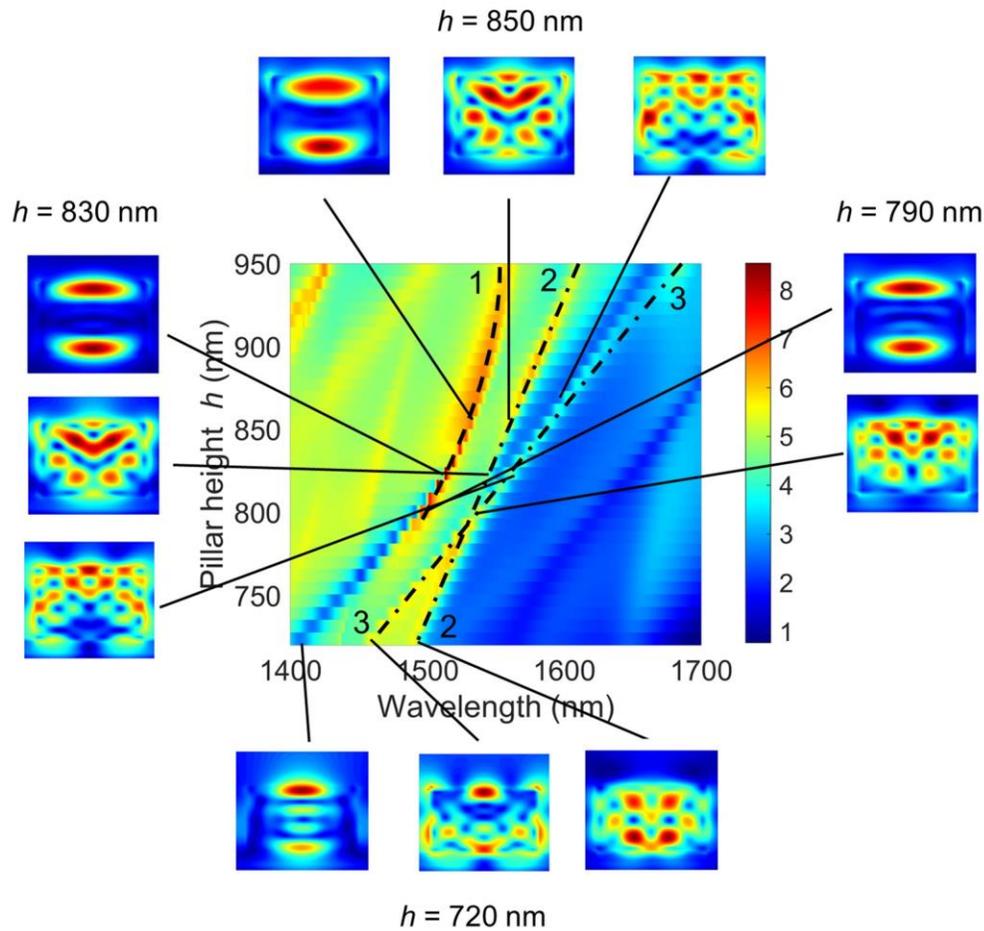

Fig. S12. Scattering cross section of a single a-Si pillar on an $SiO_2$ substrate as a function of wavelength and pillar height (the same as Fig. S11a). We plot the spatial distribution of the electric field amplitude $|E|$ in the a-Si pillar for the pillar heights of $h$ = 720 nm, $h$ = 790 nm, $h$ = 830 nm, and $h$ = 850 nm. The electric field is plotted in the x-z plane, which passes through the center of the pillar. The solid black lines point towards the scattering peaks at which the field profiles have been simulated.

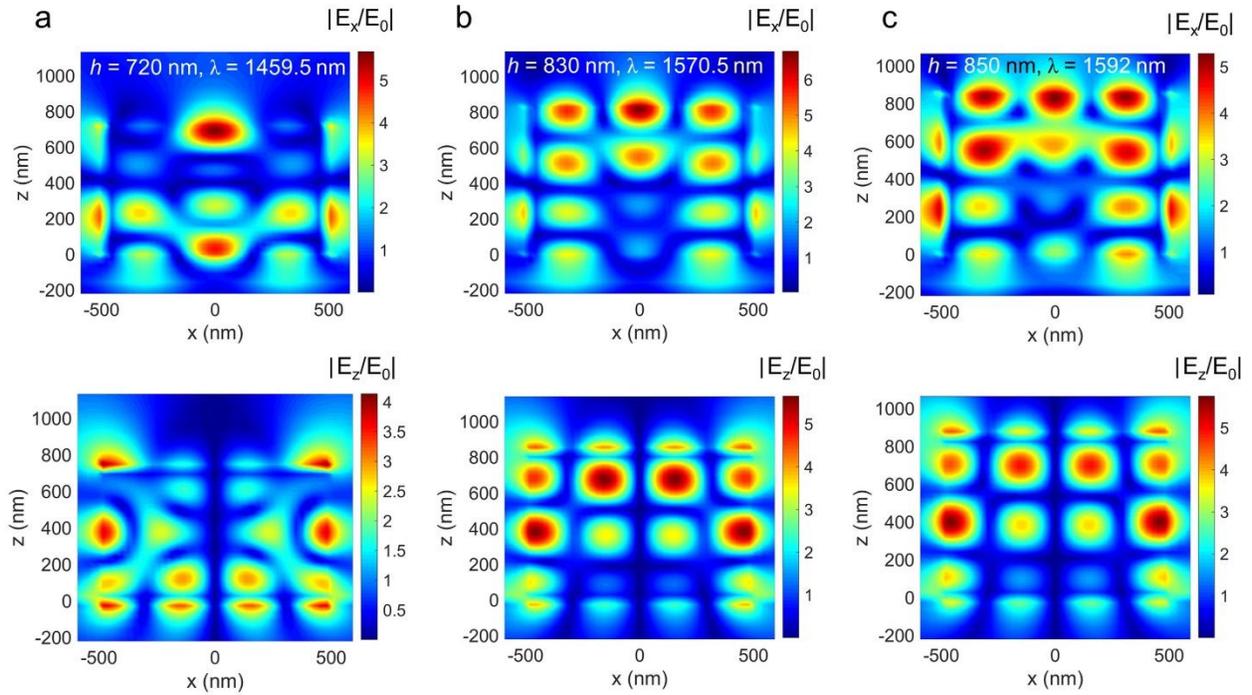

Fig. S13. Spatial distribution of the *x*-component (top panel) and *y*-component (bottom panel) of the electric field **E** in the *x-z* plane, which passes through the center of an isolated a-Si pillar on an SiO$_2$ substrate. The *x*- and *z*- components of the electric field are plotted along the *modal line 3* in Fig. S10. Geometrical parameters of the pillar are the same as in Fig. S10. The a-Si pillar height and the operating wavelength are marked at the top of each column. We observe that the *x*- and *z*- components of the electric field gradually change as we increase the pillar height.

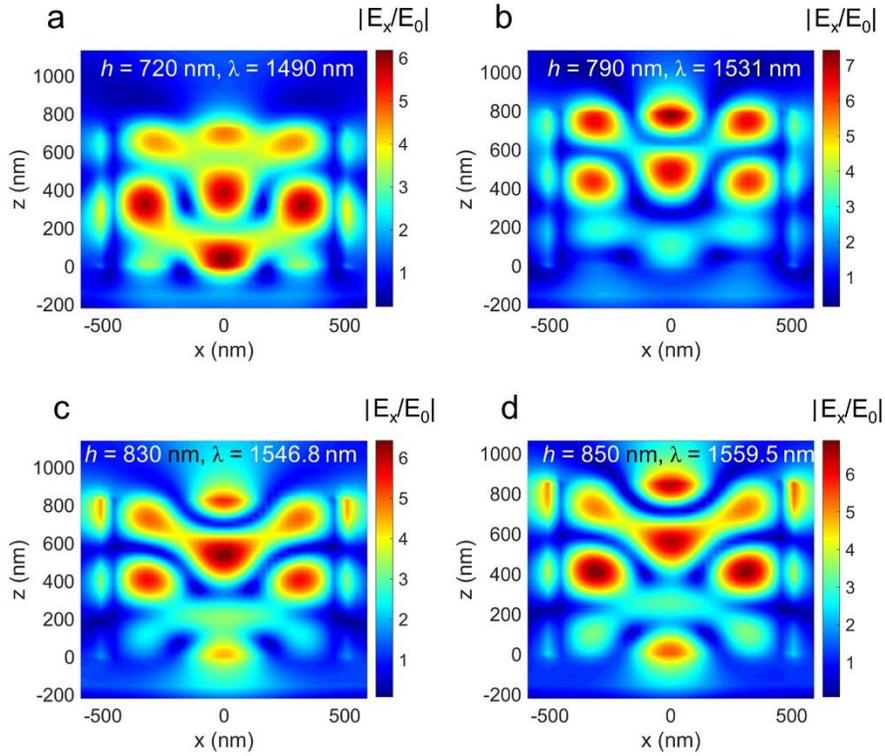

Fig. S14. Spatial distribution of the *x*-component of the electric field **E** in the *x-z* plane, which passes through the center of an isolated a-Si pillar on an SiO$_2$ substrate. The *x*-component of the electric field is plotted along the *modal line 2* in Fig. S10. Geometrical parameters of the pillar are the same as in Fig. S10. The a-Si pillar height and the operating wavelength are marked at the top of each column. We observe that the *x*-components of the electric field gradually changes as we increase the pillar height.

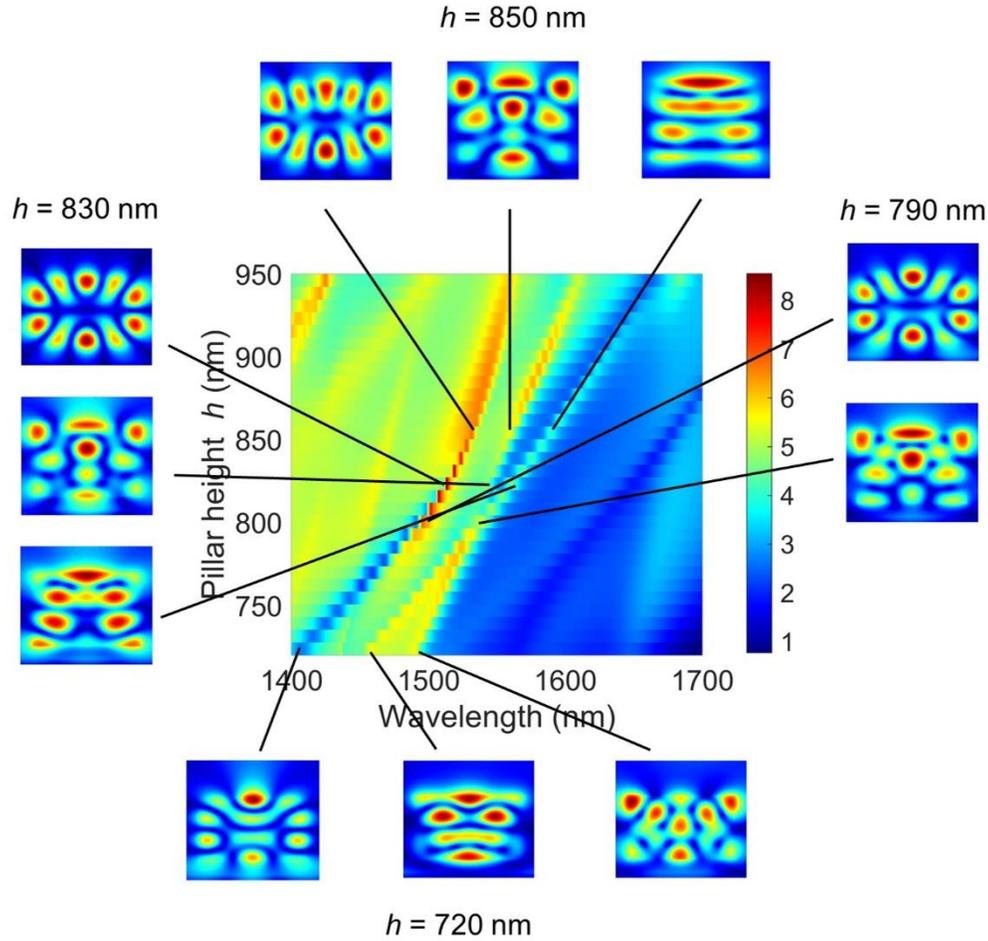

Fig. S15. Scattering cross section of a single a-Si pillar on an $SiO_2$ substrate as a function of wavelength and pillar height (the same as Fig. S11a). We plot the spatial distribution of the electric field amplitude |E| in the a-Si pillar for the pillar heights of $h$ = 720 nm, $h$ = 790 nm, $h$ = 830 nm, and $h$ = 850 nm. The electric field is plotted in the *y-z* plane, which passes through the center of the pillar. The solid black lines point towards the scattering peaks at which the field profiles have been simulated.

To understand the nature of the considered high-Q resonance and its relation to the high-Q resonance observed in the case of an array, we display the spatial distribution of the electric field amplitude in the *y-z* cross-section of the resonator (Fig. S15). As seen in Fig. S15, the spatial distribution of the electric field amplitude of the high-Q mode in the *y-z* plane (for example, at $h$ = 830 nm) is identical to the spatial distribution of the electric field observed in the case of the pillar array (see Fig. 1e).

## 6. Controlling the spectral shape of the resonance

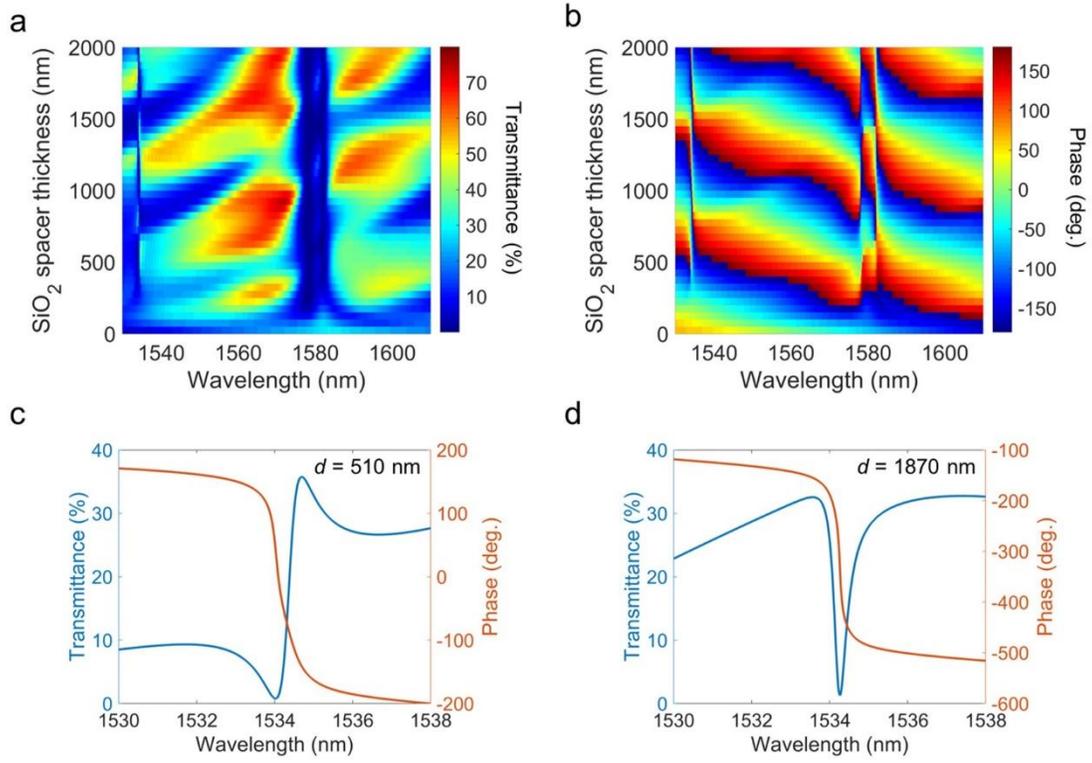

Fig. S16. Transmittance a) and phase of the transmitted light b) as a function of the wavelength and thickness of the SiO$_2$ spacer $d$ (see the schematic in Fig. 3a). The assumed geometrical parameters are the same as in Fig. 3. c) and d) show transmittance and phase spectra for the SiO$_2$ spacer thickness for $d$ = 510 nm and $d$ = 1870 nm.

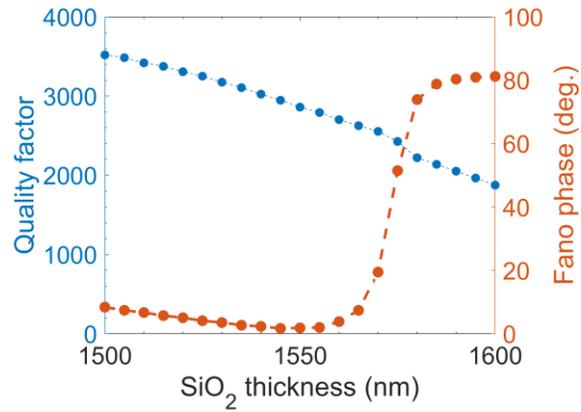

Fig. S17. Quality factor and Fano phase of the high-Q resonance as a function of the $SiO_2$ thickness *d* (for schematic see Fig. 3a). Here, the considered structure and geometrical parameters are identical to the one if Fig. 3. Figure 3c shows that around the $SiO_2$ spacer thickness *d* = 1550 nm, the Fano phase varies abruptly as a function of the $SiO_2$ spacer thickness. Here, we sample the $SiO_2$ spacer thickness more finely to accurately capture the details of the variation of the Fano phase.

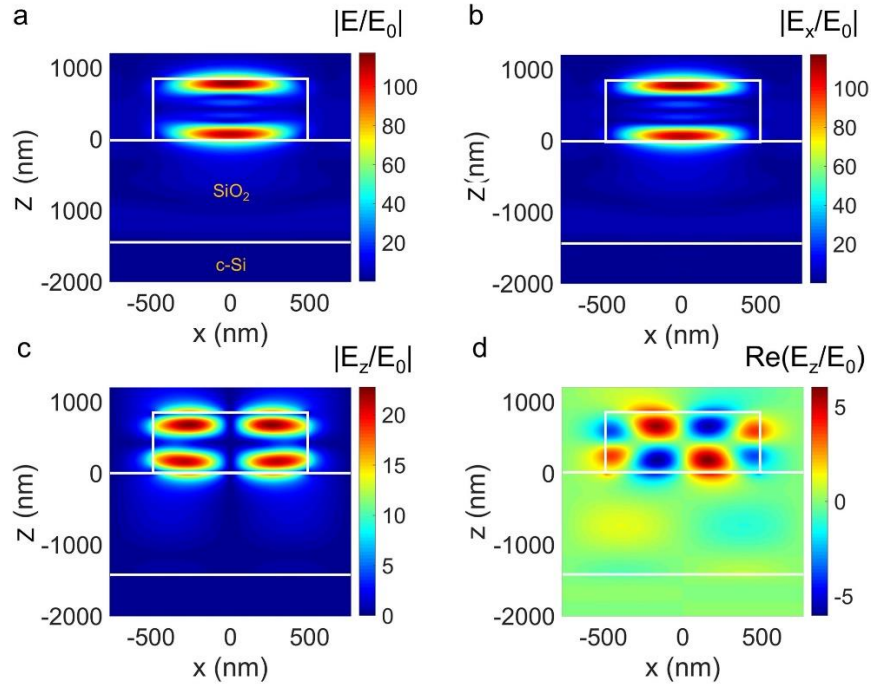

Fig. S18. Spatial distribution of the electric field amplitude in the metasurface unit cell depicted in Fig. 3a. The geometrical parameters of the metasurface unit cell are as follows: the length and width of the pillar is $l = w = 963$ nm, the height of the pillar is $h = 845$ nm. The assumed period values are $P_x = 1520$ nm and $P_y = 1425$ nm. The thickness of the $SiO_2$ spacer is $d = 1450$ nm. The spatial distribution of the electric field is plotted in the *x-y* plane. In a), we plot the absolute value of the electric field $|E/E_0|$. In b) and c), we plot the spatial distributions of *x*- and *z*-components of the electric field, respectively. In d), we plot the real part of the *z*-component of the electric field. We observe that both *x*- and *z*-components of the electric filed adopt non-zero values in the $SiO_2$ spacer.

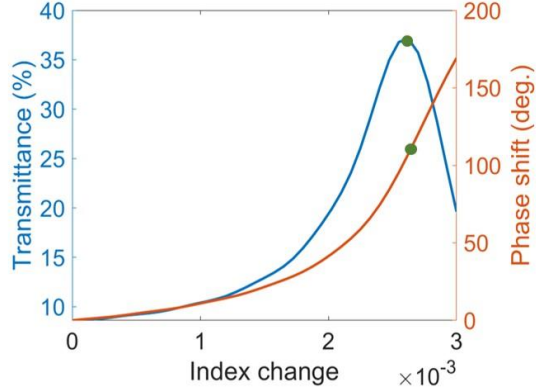

Fig. S19. Transmittance and phase shift as a function of the a-Si index change achieved by using a high-Q resonance shown in Fig. 3b. The wavelength is fixed at an operating wavelength λ = 1535.44 nm at which we have been able to fully suppress the first diffraction order (Fig. 3e). At an operating wavelength and for an index change of Δn = 0.0026, the phase shift extracted from a periodic array calculation is 107°.

7. **Realistic electrical addressing architectures**

In the present section, we describe how the proposed metasurface design can be modified to enable thermo-optic control of the wavefront of the transmitted light. The first variant of the proposed metasurface unit cell features a-Si pillars, which are connected via an a-Si bar (for a schematic of the unit cell see Fig. S20a). In Fig. S20a, the whole a-Si layer is lightly doped and is placed on an $SiO_2$ substrate so that the complex refractive index of a-Si now reads as $n$ = 3.734+0.0013 i. We can actively control the temperature of the metasurface pixel by biasing the metasurface unit cell at the edge and flowing current in the $y$ direction. the geometrical parameters of the proposed metasurface unit cell are as follows: δ = 82 nm, $w = l_1$ =963 nm, $h$ = 850 nm, $P_x = P_y$ = 1500 nm. The inset of Fig. S20b shows a schematic of one period of a two-level phase grating, which is utilized to theoretically demonstrate switchable diffraction. Within a grating period, the assumed index change between neighboring a-Si pillars is Δn=0.006 (Figs. S20b and S20a). The switchable diffraction is observed both in the case when the polarization of the electric field is perpendicular and parallel to the a-Si bars (Figs. S20b and S20c, respectively). In Fig. S20b, the operating wavelength is 1586.5 nm, and the overall transmittance at the operating wavelength is $T$ = 1.2%. In Fig. S20c, the operating wavelength is 1583.2 nm, and the overall transmittance at the operating wavelength is $T$=0.3%. Note that to observe the diffractive switching, we have utilized the lower-Q mode described in Fig. 3 of the main manuscript. When using the configuration shown in Fig. S20a, we are not able to demonstrate diffractive beam switching using the higher-Q mode (see Fig. 1 of the main manuscript) since the spectral characteristics of the high-Q mode are strongly affected by introduced optical losses. Namely, in the lossy structure (Fig. S20a), we do not observe large spectral variation of the phase of the transmitted light in the case of no applied current.

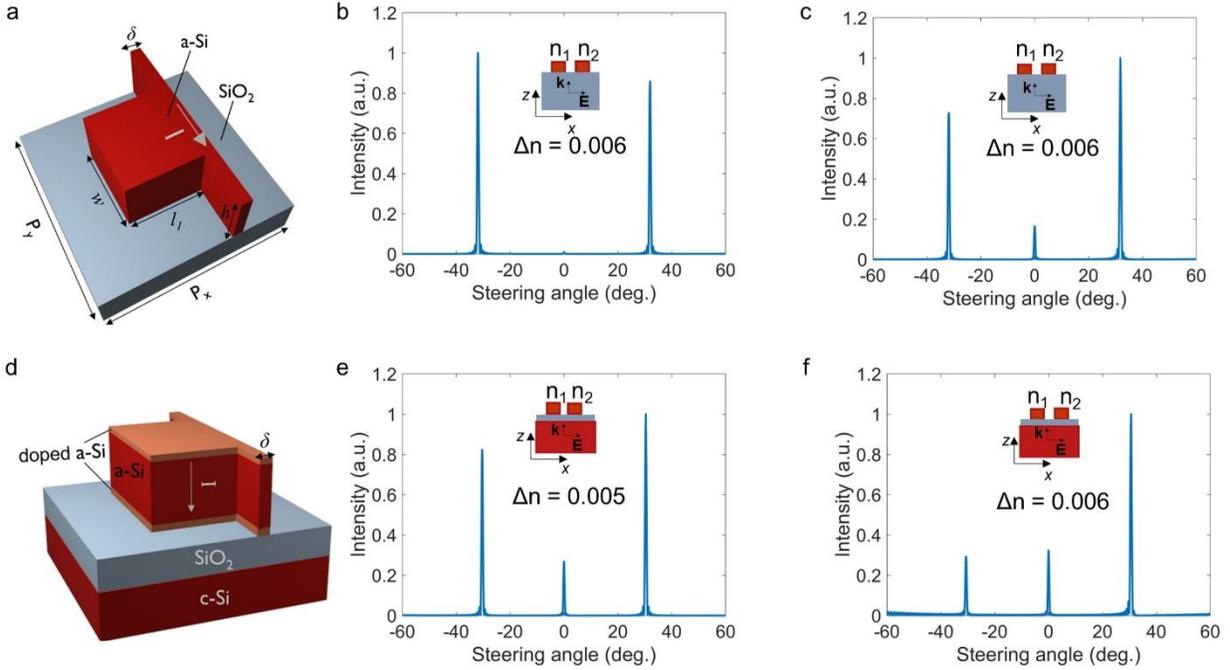

Fig. S20. Dynamic beam switching with realistic interconnect architectures. a) Schematic of the proposed metasurface unit cell. The square a-Si pillars connected via a-Si bars. In a) the a-Si is lightly doped. b) and c) plot the electric field intensity in the far field as function of the steering angle for a two-level phase grating, which uses the unit cell shown in a). The a-Si refractive index difference between neighboring metasurface elements is $\Delta n$ = 0.006. In b), the incoming plane wave is x-polarized while in c) the incoming electric field is y-polarized. d) Schematic of a metasurface unit cell in which the top and bottom 50 nm of a-Si are doped. The thickness of the $SiO_2$ layer is 1440 nm. e) and f) plot the electric field intensity in the far field as function of the steering angle for a two-level phase grating, which uses the unit cell shown in d). In e) the a-Si refractive index difference between neighboring metasurface elements is $\Delta n$ = 0.005, and the width of the a-Si bar is $\delta$ = 40 nm. In f) the a-Si refractive index difference between neighboring metasurface elements is $\Delta n$ = 0.006, and the width of the a-Si bar is $\delta$ = 82 nm.

   To enable thermo-optic switching using also higher-Q mode we modify our metasurface unit cell design. In the modified unit cell design, only top and bottom 50 nm layers of the a-Si are doped while the core section of a-Si layer is practically undoped (or very lightly doped). In this implementation, the top and bottom electrodes are biased with respect to each other, and current flows in the vertical direction. We have also slightly modified the geometrical parameters of the metasurface unit cell. In S13d, $P_x$ = 1520 nm and $P_y$ = 1425 nm. The height of the a-Si and the a-Si bar is $h$ = 845 nm, and the thickness of the $SiO_2$ spacer is 1440 nm. The width and length of the a-Si pillar are taken to be $w = l_1$ = 963 nm. The insets of Figs. S20e and S20f show schematics of one period of a two-level phase grating, which is utilized to theoretically demonstrate switchable diffraction. Utilizing the higher-Q mode, we have been able to observe switchable diffraction assuming that the refractive index difference between neighboring a-Si pillars in a grating period is Δ$n$ = 0.005, and the width of the a-Si bar of δ = 40 nm. In Fig. S20e, the polarization direction of the electric field of the incoming light is perpendicular to the a-Si bars. In Fig. S20e, the operating wavelength is 1536.15 nm, while the overall transmittance at the operating wavelength is $T$ = 6.8 %.

Interestingly, using the two-level phase grating, we have been able to observe highly asymmetric diffraction pattern originating from geometrical asymmetry of the proposed metasurface unit cell (Fig. S20f). In Fig. S20f, the refractive index difference between neighboring a-Si pillars in a grating period is Δ$n$ = 0.006. The width of the a-Si bar is taken to be δ = 82 nm. the operating wavelength of the utilized lower-Q mode is 1548.4 nm, and the overall transmittance at the operating wavelength is $T$ = 1 %.

Finally, we consider the case when the metasurface unit cells are placed on an $SiO_2$ pedestal and the a-Si pillars are connected in series vis a-Si bars. In this case, for the pillar height of $h$ = 880 nm, the spectral and phase signatures of the high-Q mode are not observed.

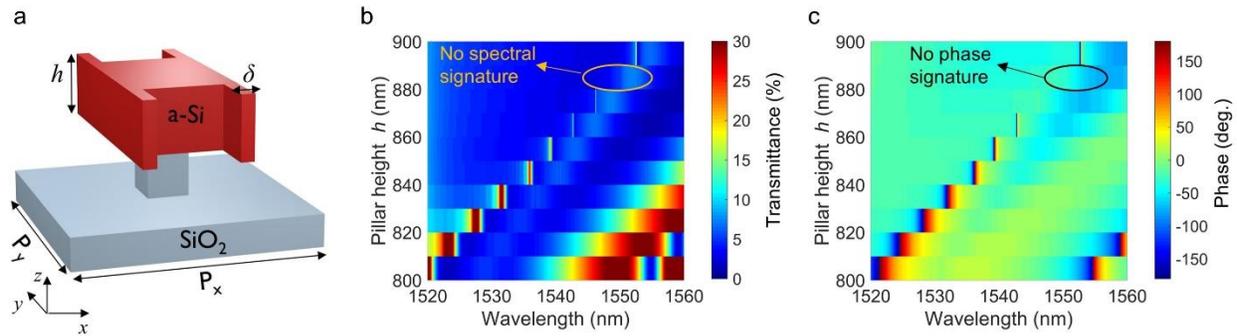

Fig. S21. Dependence of the transmittance and phase on the height of the pillar and bars. a) Schematic of the metasurface unit cell. In the schematic, $P_x$ = 1520 nm, $P_y$ =1 425 nm, δ = 82 nm. The height of the $SiO_2$ pedestal is 382 nm. The length of the pedestal is 200 nm, the width of the pedestal is also 200 nm. Note that the metasurface considered here does not include a Si substrate. In b), we plot transmittance as a function of wavelength and the a-Si pillar height. In c), we the phase of the transmitted light as a function of wavelength and the a-Si pillar height. When changing the height of the a-Si pillar, we simultaneously change the height of the a-Si bars. As seen in b) and d), at the pillar height of $h$ = 880 nm, the spectral and phase signature of the resonance disappears.

The metasurface structure considered in the present section is shown in Figs. 5a and 5b. The geometrical parameters of the considered structure are specified in the caption of Fig. 5.

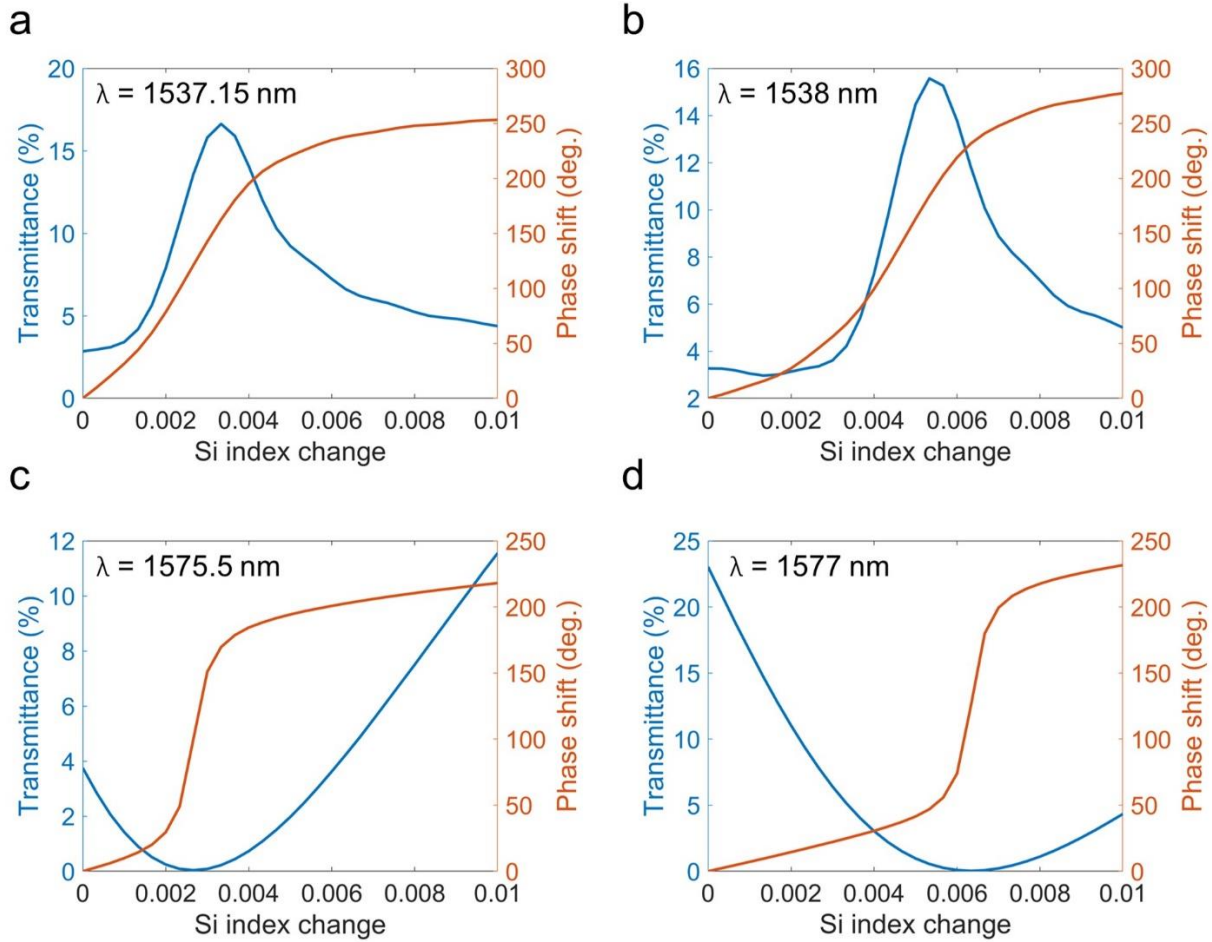

Fig. S22. Transmittance and phase shift as function of the a-Si index change for the metasurface with electrodes shown in Figs. 5a and 5b. The incoming light is *x* polarized. a) and b) correspond to the high-Q mode while c) and d) correspond to the lower-Q mode. The maximal phase shift enabled by the high-Q mode is 277° while the lower-Q mode enables a phase shift of 230° when the refractive index of a-Si is changed by Δ*n* = 0.01.

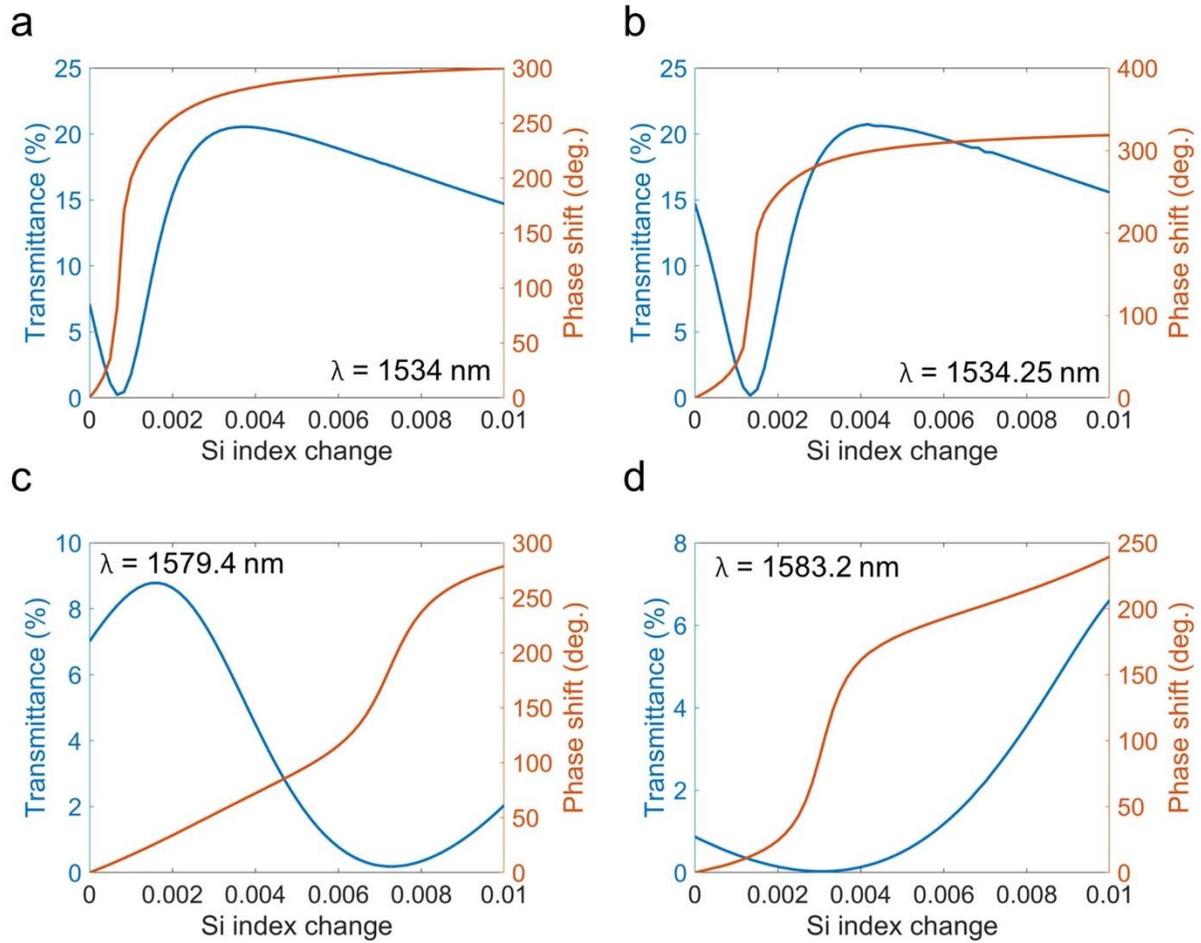

Fig. S23. Transmittance and phase shift as function of the a-Si index change for the metasurface with electrodes shown in Figs. 5a and 5b. The incoming light is *y* polarized. a) and b) correspond to the high-Q mode while c) and d) correspond to the lower-Q mode. The maximal phase shift enabled by the high-Q mode is 320° while the lower-Q mode enables a phase shift of 275° when the refractive index of a-Si is changed by Δ*n* = 0.01.

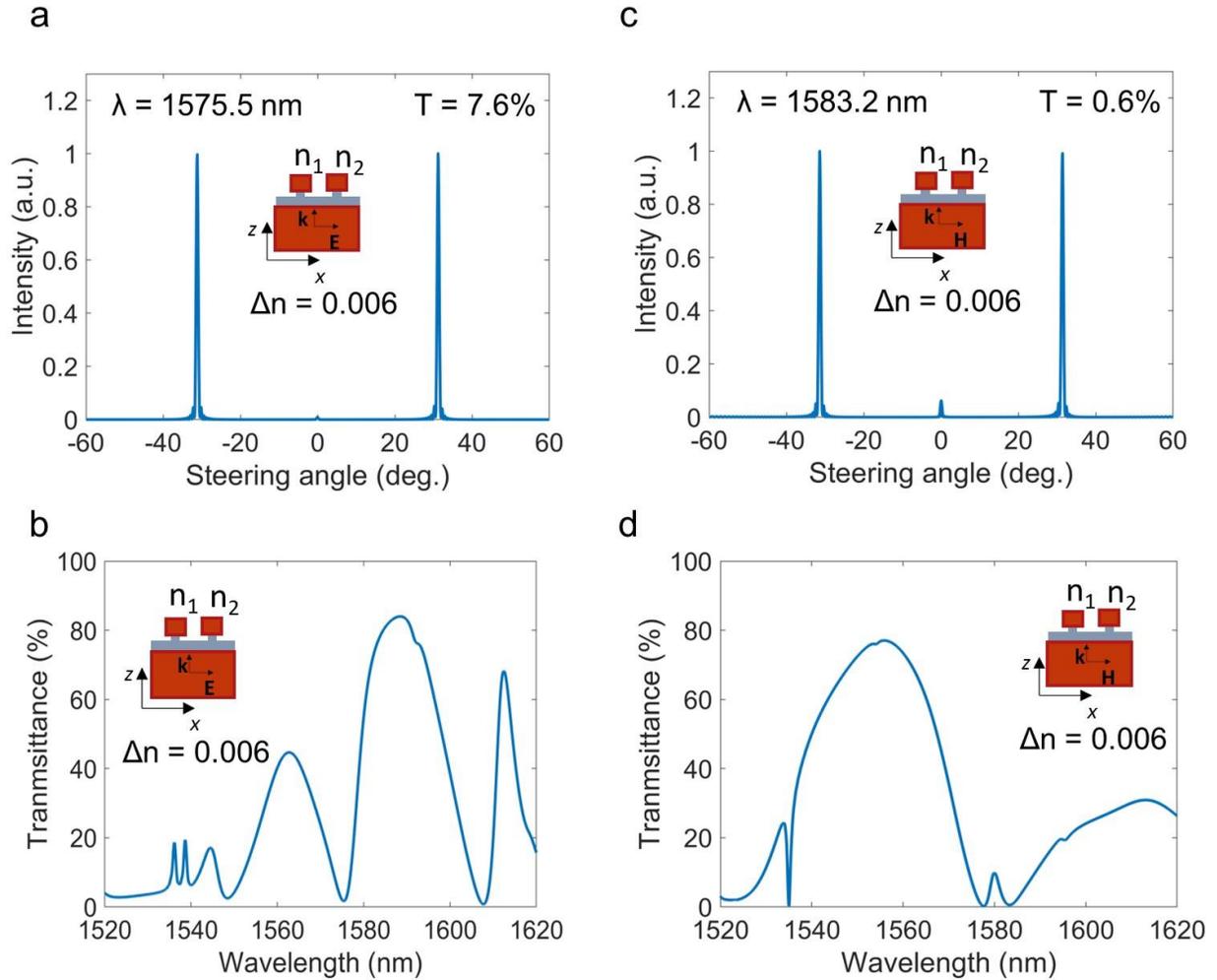

Fig. S24. Thermo-optic beam switching using the lower-Q mode. The metasurface unit cell is given by the schematic shown in Figure 5a. We analyze the optical performance of a two-level phase grating. The unit cell of the grating is shown in the inset of a). The index change between neighboring a-Si pillars is Δ$n$ = 0.006. a) The intensity of the electric field in the far field when the incoming plane wave is x polarized, the operating wavelength is λ = 1575.5 nm, and the overall transmittance is T = 7.6 %. In b), the overall transmittance spectrum of the two-level grating studied in a). In c), The intensity of the electric field in the far field when the incoming plane wave is y polarized. In c), the operating wavelength is λ = 1583.2 nm, and the overall transmittance is T = 0.6 %. In d), the overall transmittance spectrum of the two-level grating studied in c).

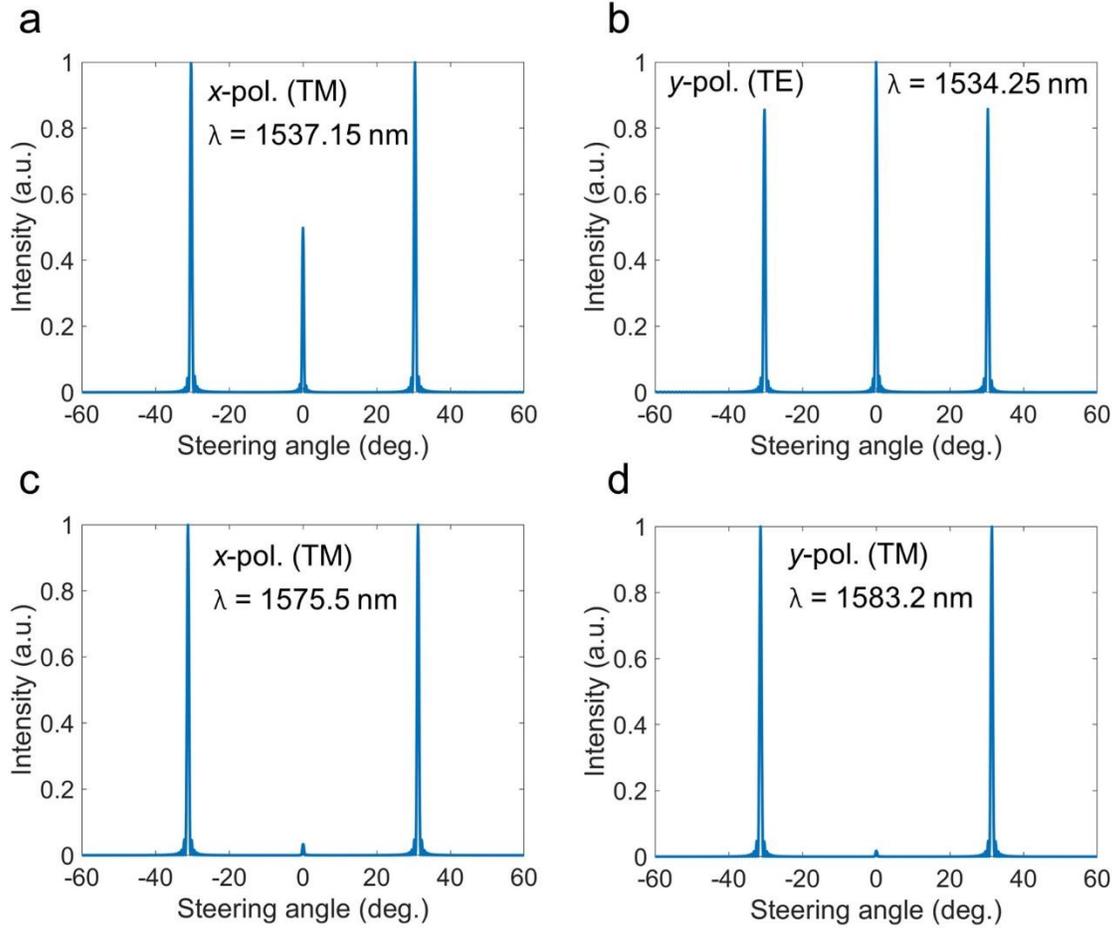

Fig. S25. Analytical array factor calculations for a two-level phase grating [3]. The plots calculated intensity of the electric field in the far field as a function of the steering angle in the case of a two-level phase grating. a) and b) correspond to the case of the high-Q mode. In a) and b), we assume that the refractive index difference between neighboring metasurface pixels is 0.0026, which is the value chosen in Fig. 5. In a) and b), the values of the electric field amplitudes and relative phases in a grating period are constructed based on Figs. S22a and S23b. c) and d) correspond to the case of the lower-Q mode. In c) and d), we assume that the refractive index difference between neighboring metasurface pixels is 0.006, which is the value chosen in Fig. S24. In c) and d), the values of the electric field amplitudes and relative phases in a grating period are constructed based on Figs. S22c and S23d. We observe that, in the case of the high-Q mode, the array level analytical calculations yield significantly different results as compared with the case of full wave simulations providing evidence of near-field coupling between neighboring metasurface pixels [cf. c) and Fig. 5c as well as d) and Fig. 5e]. On the other hand, in the case of the lower-Q mode, the array factor calculation results and full wave simulations yield similar results (cf. c) and Fig. S24a as well as d) and Fig. S24).

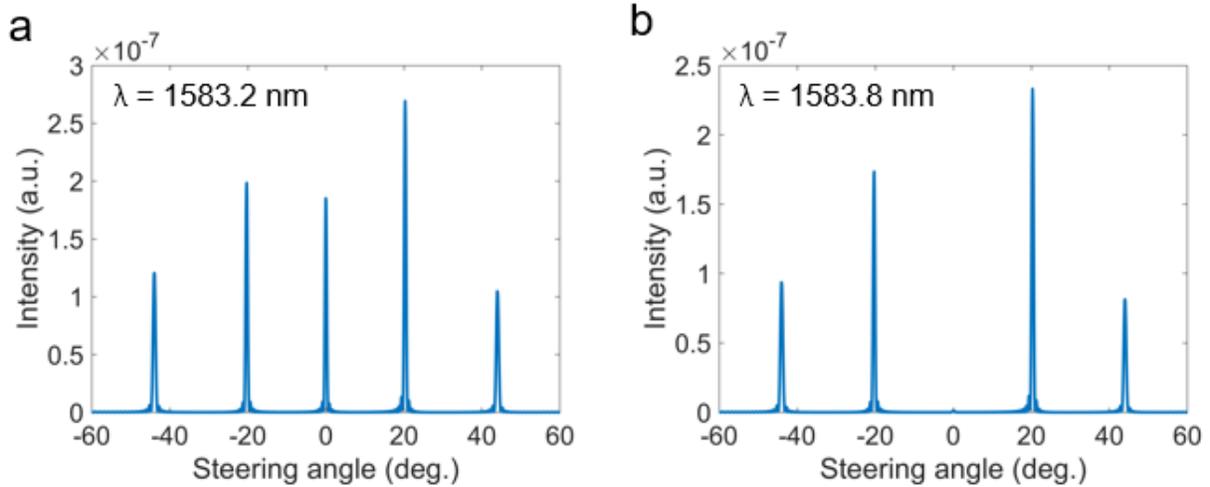

Fig. S26. Intensity of the electric filed in the far field as a function of the steering angle in the cases of a three-level phase grating. The incoming light is *y* polarized. The three-level phase grating is generated using the simulated phase shift data for an operating wavelength of λ = 1583.2 nm (Fig. S23), and the phase shift values in the grating period are given as (0°, 120°, 240°). Within a grating period, the values of the real part of the refractive index are taken as (3.734, 3.734+0.003333, 3.734+0.01). a) plots the intensity of the electric field in the far field at an operating wavelength of λ = 1583.2 nm. In b), we keep the same spatial distribution of the real part of the refractive index as in a), but the operating wavelength is now taken as λ = 1583.8 nm. In a) we observe the target steered beam at a steering angle of 20.3°. Additionally, we observe a number of spurious diffraction orders. By changing the operating wavelength to λ = 1583.8 nm, we are able to fully suppress the zeroth diffraction order, but the other spurious diffraction orders are still observed.

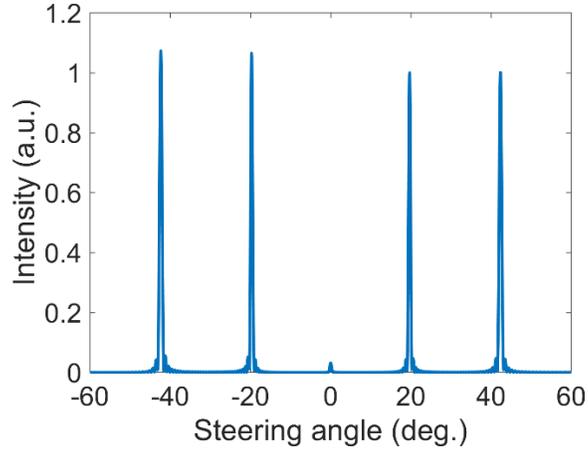

Fig. S27. Intensity of the electric filed in the far field as a function of the steering angle in the cases of a three-level phase grating at an operating wavelength of λ = 1535.81 nm, which is a resonant wavelength for a high-Q mode. The incoming light is *y* polarized. Within a grating period, the values of the real part of the refractive index are taken as (3.734, 3.734+0.003333, 3.734+0.01). By appropriately choosing the operating wavelength, we have been able to suppress the zeroth diffraction order.

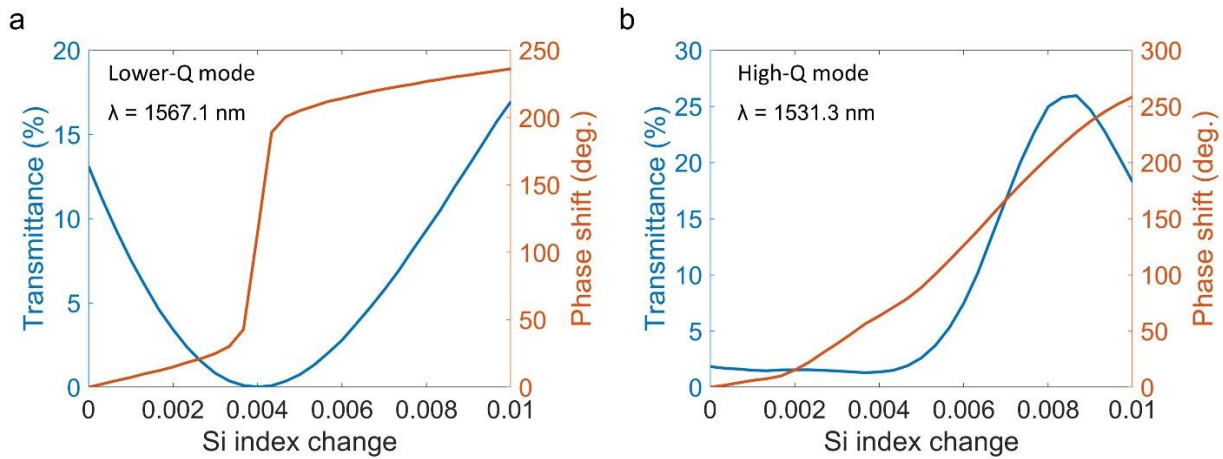

Fig. S28. Transmittance and phase shift as function of the a-Si index change for an optimized metasurface with electrodes shown in Figs. 5a and 5b. The incoming light is *x*-polarized. a) corresponds to the lower-Q mode and b) correspond to the high-Q mode. The insets indicate the operating wavelengths for which the diffraction patterns shown in Fig. 7 are observed. At an operating wavelength of λ = 1567.1 nm, the maximal phase shift enabled by the lower-Q mode is 236.2°. In the case of the high-Q mode, the maximal phase shift at an operating wavelength of λ = 1531.3 nm is 265°.

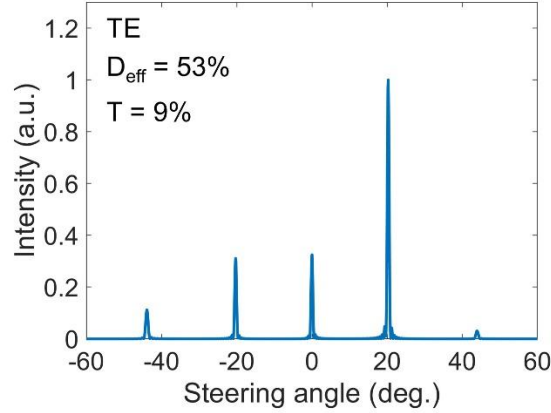

Fig. S29. Thermo-optic three-level phase grating with a realistic interconnect architecture and optimized geometry. The incoming plane wave is TE-polarized. The figure plots the intensity of the electric field in the far field for the case of the lower-Q mode at an operating wavelength of λ=1581.5 nm. The employed optimization procedure has yielded the following geometrical parameters for the structure shown in Figs. 5a-b: $P_x$ = 1520 nm, $P_y$ = 1100 nm, and $h$ = 851.327 nm, and the resulting diffraction efficiency $D_{eff}$=53%. The real part of the complex refractive index for each metasurface element within the grating period is given as follows: $n_1$ = 3.734, $n_2$ = 3.74144, $n_3$ = 3.74318.

## 8. Metasurfaces with a finite number of metasurface elements

We consider an *m* x *m* array of metasurface elements and study how the quality factor of the mode supported by the finite array increases with *m* (see Fig. S30). To calculate the quality factor, we calculate the scattering cross-section of the finite array and fit it to the Fano formula (see Section 1). When studying a finite metasurface array, the assumed geometrical parameters are the same as in Fig. 1 of the main manuscript: $l$ = $w$ = 963 nm and $P_x$ = $P_y$ = 1425 nm. To reduce the simulation time, the assumed mesh in the *z*-direction is 10 nm while the mesh in the *x*- and *y*-directions is set to 20 nm. When assuming periodic boundary conditions in the x- and y-directions, the quality factor of the supported mode is 8500. Note that this quality factor value (8500) is lower as compared with the one reported in Figs. 1 and 2 of the manuscript (9800). This difference in the observed quality factors is due to the fact the in the simulations shown in Fig. 1, the mesh in the *z*-direction is taken to be 5 nm while the mesh in the *x*- and *y*-directions is still set to 20 nm. As seen in Fig. S30, the quality factor of the supported mode monotonously increases when increasing the number of metasurface elements in the array. For a 12x12 array, the quality factor is 6000.

Next, we assume that the number of the metasurface elements is finite in the direction *parallel* to the incoming electric field while in the direction *perpendicular* to the electric field, the periodic boundary condition is assumed (see the inset of Fig. S30b). The described simulation setup enables reducing the simulation time while accessing the quality factor of the metasurface for a larger number of metasurface elements $n_x$. As seen in Fig. S30b, when the number of the metasurface elements in the direction parallel to the electric field nx is $n_x$ = 10, the quality factor of the metasurfaces is ~7000. While when $n_x$ = 20, the quality factor of the metasurface is slightly below 8000.

In Fig. S30c, we assume that the number of the metasurface elements is finite in the direction *perpendicular* to the incoming electric field while in the direction *parallel* to the electric field, the periodic boundary condition is assumed. In this case, when $n_x$ = 10, the quality fact of the metasurface is ~6000. Note though that in both cases where the metasurface is finite in the direction *perpendicular* or *parallel* to the electric field, for $n_x$ = 20, the quality factor of the metasurface is ~8000.

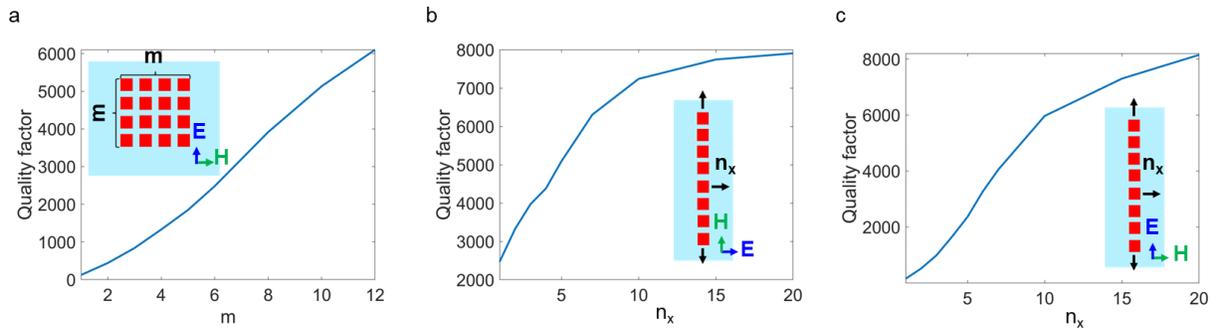

Fig. S30. Quality factors of metasurfaces with a finite number of elements. a) Quality factor of an $m \times m$ metasurface array as a function of m. b) Quality factor of the metasurface, which is finite in the direction *parallel* to the incoming electric field as a function of the number of the metasurface elements in the mentioned direction $n_x$. In b), in the direction perpendicular to the electric field, the periodic boundary condition is assumed. c) Quality factor of the metasurface, which is finite in the direction *perpendicular* to the incoming electric field as a function of the number of the metasurface elements in the mentioned direction $n_x$. In c), in the direction parallel to the electric field, the periodic boundary condition is assumed.

9. **High-efficiency reflective high-Q metasurfaces.**

By adding an Au back reflector to the designed transmissive metasurface we can attain high-efficiency high-Q reflective metasurfaces. The unit cell of the proposed metasurface design is shown in Fig. S31a. The metasurface unit cell consists of an Au back reflector, followed by an $SiO_2$ spacer on top of which we place an a-Si pillar. By appropriately choosing the thickness of the $SiO_2$ spacer we obtain high-efficiency high-Q reflective metasurfaces.

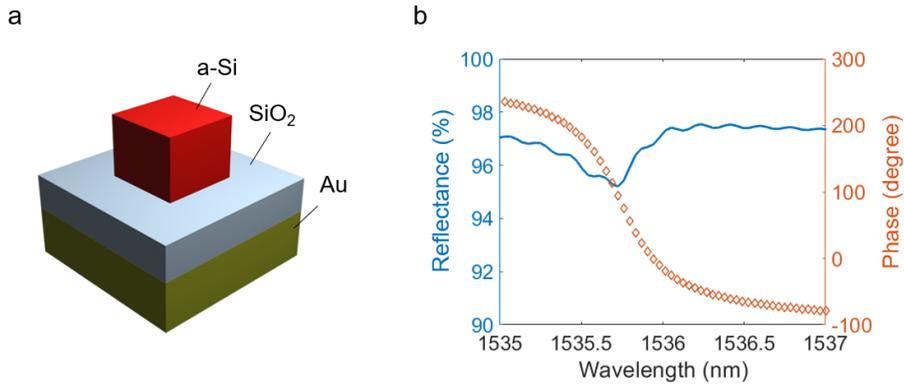

Fig. S31. High-efficiency transmissive metasurfaces. a) Schematic of the unit cell of the high efficiency transmissive metasurface. In a), the length and width of the a-Si pillar are taken to be $l = w = 963$ nm. The metasurface period is $P_x = P_y = 1425$ nm. the thickness of the SiO$_2$ spacer is 516 nm while Au is optically thick. b) Phase and reflectance spectra for the designed metasurface.